\documentclass[pra, reprint, superscriptaddress]{revtex4-2}
\usepackage[dvipsnames]{xcolor}
\usepackage[sort&compress]{natbib}
\usepackage{mathtools}
\usepackage{amsfonts}
\usepackage{physics}
\usepackage[a4paper]{geometry}
\newgeometry{top=1.5cm,right=1.5cm,left=1.5cm,bottom=1.5cm}
\usepackage[colorlinks=true, urlcolor=blue, pdfborder={0 0 0}]{hyperref}
\hypersetup{linkcolor=blue,citecolor=BrickRed}
\usepackage{parskip}
\begin{document}

\preprint{APS/123-QED}

\title{Artificial Neural Network Syndrome Decoding on IBM Quantum Processors}

\author{Brhyeton Hall}
\email{brhyeton@unimelb.edu.au}
\affiliation{%
School of Physics, The University of Melbourne, Parkville, 3010, Victoria, Australia
}%

\author{Spiro Gicev}
\email{gicevs@unimelb.edu.au}
\affiliation{%
School of Physics, The University of Melbourne, Parkville, 3010, Victoria, Australia
}%

\author{Muhammad Usman}
\email{musman@unimelb.edu.au}
\affiliation{%
School of Physics, The University of Melbourne, Parkville, 3010, Victoria, Australia
}%
\affiliation{%
Data61, CSIRO, Research Way, Clayton, 3168, Victoria, Australia
}%

\begin{abstract}
\noindent

Syndrome decoding is an integral but computationally demanding step in the implementation of quantum error correction for fault-tolerant quantum computing. Here, we report the development and benchmarking of Artificial Neural Network (ANN) decoding on IBM Quantum Processors. We demonstrate that ANNs can efficiently decode syndrome measurement data from heavy-hexagonal code architecture and apply appropriate corrections to facilitate error protection. The current physical error rates of IBM devices are above the code's threshold and restrict the scope of our ANN decoder for logical error rate suppression. However, our work confirms the applicability of ANN decoding methods of syndrome data retrieved from experimental devices and establishes machine learning as a promising pathway for quantum error correction when quantum devices with below threshold error rates become available in the near future.       
\end{abstract}

\maketitle

\begin{figure*}[t]
    \centering
    \includegraphics[width=\linewidth]{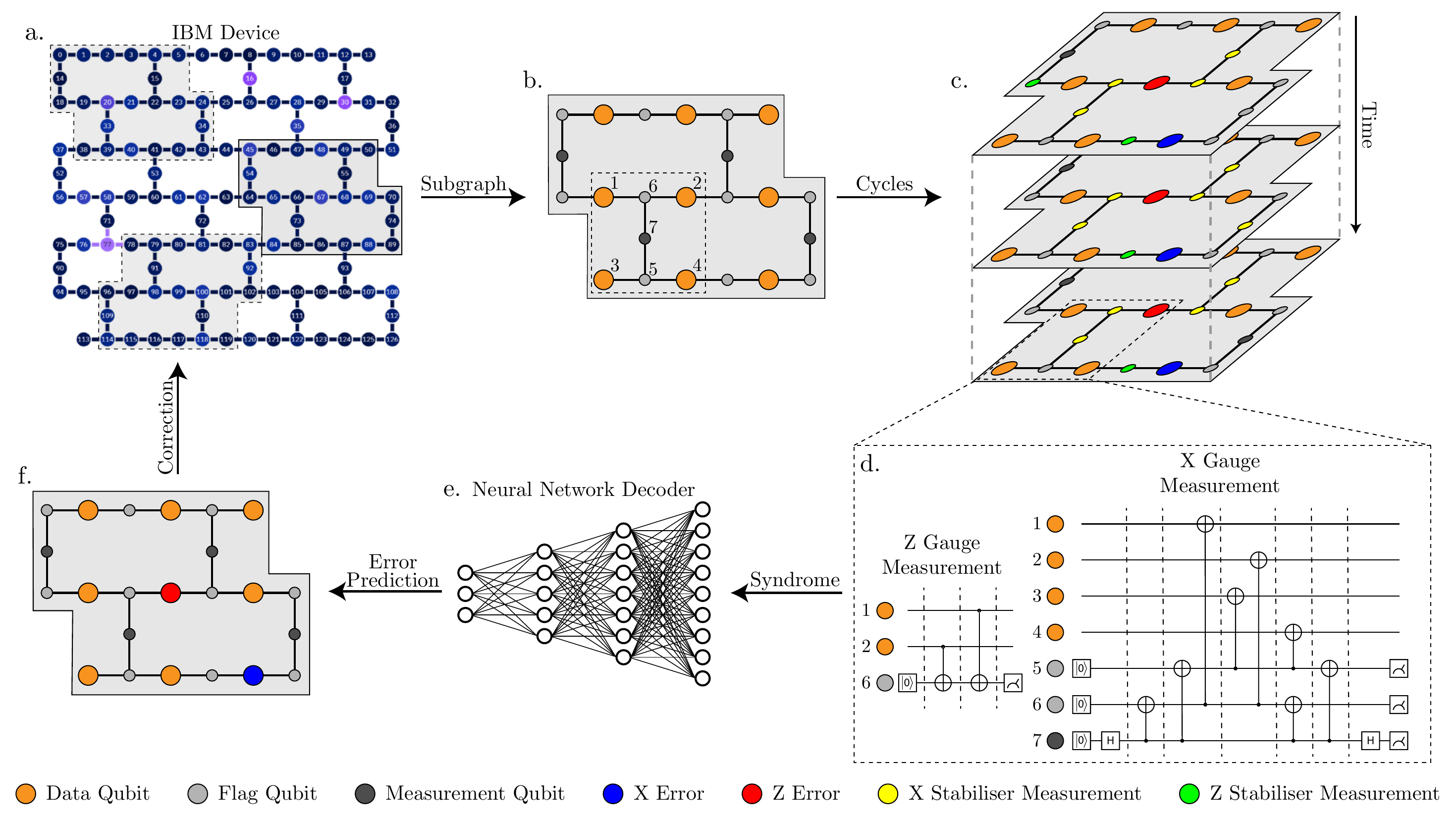}
    \caption{\textbf{Neural Network Decoder Framework.} \textbf{a.} shows the lattice connectivity of qubits of a 127 qubit device developed by IBM with colour range denoting error probabilities associated with single and two qubit gates; lighter being more error prone. The shaded section represents a subsection of this device where the average error rate is lowest, in a region which supports a $d=3$ HH error correction code. Dotted outlines indicate some other possible subgraph locations. \textbf{b.} show the qubits of a HH code, with orange circles representing the data qubits and light/dark grey circles representing the ancillary flag and measurement qubits respectively. Connecting lines represent the connectivity of two-qubit gates within the lattice. \textbf{c.} shows multiple cycles of the HH error syndrome measurement in the presence of circuit noise, including state preparation, readout and idle qubit errors. \textbf{d.} illustrates the circuits for $X$ and $Z$ gauge operator measurement of the HH code. \textbf{e.} represents an ANN based syndrome decoder as developed in this work. A large input layer takes the measurements over $d$ cycles, and it linearly decreases over four layers to an output which is the size of the number of data qubits. \textbf{f.} shows a possible correction being sampled from the prediction give by the ANN based syndrome decoder. The appropriate correction is then applied to the IBM device.}
    \label{fig:fig1}
\end{figure*}

The development of quantum processors has made remarkable progress over the last few years with quantum devices consisting of more than 100 qubits currently accessible from multiple developers \cite{collins_ibm_2022, madsen_quantum_2022, barnes_assembly_2022}. 
In principle, 100 qubits could allow computations intractable on classical supercomputers, however, the computational capabilities of the current generation of quantum processors are limited by high levels of physical noise \cite{beverland_assessing_2022}. 
Several studies have implemented and tested error mitigation strategies to suppress the detrimental impact of noise with varying levels of success \cite{kim_evidence_2023, endo_practical_2018, strikis_learning-based_2021, bravyi_mitigating_2021}. 
Ultimately, the full power of quantum computers can only be unleashed when Quantum Error Correction (QEC) techniques are implemented. 
These will allow efficient and scalable detection and correction of errors in quantum circuits, leading to fault-tolerant quantum computations \cite{shor_scheme_1995, kitaev_fault-tolerant_2003, aharonov_fault-tolerant_2008, knill_resilient_1998}. 
Over the recent decades, QEC codes have been theoretically developed to provide a means to suppress errors on logical information through the use of encoding in a larger Hilbert space \cite{gottesman_stabilizer_1997, steane_active_1997, knill_resilient_1998, kitaev_quantum_1997}. 
One of the leading QEC codes is the surface code, which offers a high logical error rate threshold based on nearest neighbour interactions between qubits on a two-dimensional lattice \cite{kitaev_fault-tolerant_2003, fowler_surface_2012}. 
The implementation of surface code based QEC requires the classical processing of syndrome data --- related to the physical error locations --- to find appropriate corrections for physical qubits.
However, this step, known as decoding, is a computationally intensive task. 
Recent work has theoretically shown that Artificial Neural Network (ANN) based decoders can facilitate fast and scalable decoding \cite{varsamopoulos_decoding_2017, meinerz_scalable_2022, varsamopoulos_comparing_2020, gicev_scalable_2023, overwater_neural-network_2022, wagner_symmetries_2020, ni_neural_2020, zhang_scalable_2023} which is crucial to prevent the accumulation of errors during any quantum computation. 
The next major milestone is to implement an ANN based syndrome decoder on quantum processors to directly benchmark their performance. 
This has only been reported by one recent paper to-date, which is based on quantum devices developed by the Google team \cite{bausch_learning_2023}.

In this work, we develop an ANN based syndrome decoder and experimentally implement it on IBM Quantum Processors. 
Further, we assess its performance through comparison against the well established graph-based Minimum Weight Perfect Matching (MWPM) technique, using PyMatching \cite{higgott_pymatching_2022}. 
Our work shows that, in principle, ANN based syndrome decoders can efficiently process syndrome measurement data from IBM devices and suggest appropriate corrections --- achieving a crucial step in the pipeline of QEC on quantum computational devices. 

Historically, the development of surface code literature has been primarily based on the square lattice arrangement of qubits \cite{kitaev_fault-tolerant_2003, dennis_topological_2002, fowler_surface_2012}, however the architecture of the IBM Quantum Processors is built on a heavy-hexagonal (HH) arrangement of qubits, as shown in Figure \ref{fig:fig1} (a). 
The motivation for such a qubit layout was to reduce the local connectivity of qubits.
This addressed the physical difficulty of controlling many connections to each qubit and aimed to reduce cross-talk noise \cite{chamberland_topological_2020}.
However, the HH format required the modification of the traditional square surface code construction to a hexagonal architecture, with ancillary qubits --- changing the underlying circuit structures for syndrome measurement.
In 2020, Chamberland \textit{et al.} laid out the foundational framework for QEC on heavy-hexagonal and heavy-square lattices of low-degree locally connected qubits \cite{chamberland_topological_2020} --- introducing the HH QEC code. 
This original HH code was optimised to minimise the number of required physical qubits by removing some ancillary qubits on the boundaries of the hexagonal lattice and maintaining a lattice connectivity of at most, three \cite{chamberland_topological_2020}.
However, IBM have developed increasingly large devices on HH lattices, without the original optimisation of boundaries \cite{nation_ibm_2021}, as shown in Figure \ref{fig:fig1} (a), as the original code layout was incompatible with being realised in the bulk of a HH lattice. 
This created a discrepancy between the HH code proposed by Chamberland \textit{et al.} and the HH layout of physical qubits in IBM devices (see Supplementary Material Section \ref{sec:adjustment} for details on the adjustment made). 
To address this disparity, we have modified the existing HH code, by adjusting the original prescription's boundaries to fit with the bulk. 
This conforms with the IBM Quantum Processor layout, which is a crucial step in the direct implementation and benchmarking of our ANN decoder on IBM devices. 
A recent work by Sundaresan \textit{et al.} has also looked at the modified HH code for distance three measurements \cite{sundaresan_demonstrating_2023}.
However, our work is distinct as we investigate HH code threshold plots and implementation comparison between distance three and five codes based on direct measurements on IBM devices.

Figure \ref{fig:fig1} (b) schematically illustrates a distance three patch of the adjusted HH code shape as described within our work, where data qubits (orange) store useful information and ancilla qubits (grey) are used to facilitate syndrome measurements. 
These measurements are used to locate errors on physical qubits within the HH lattice. 
Typically, the syndrome measurements are collected in multiple rounds before they are decoded to find appropriate corrections for data qubit errors and also in the syndrome measurement process itself. 
Figure \ref{fig:fig1} (c) schematically shows many cycles of the HH code being executed and corresponding syndromes measured for each cycle. 
The circuits that are used to measure the syndrome in both the $X$ and $Z$ basis are shown in Figure \ref{fig:fig1} (d), with the physical qubits numbered in Figure \ref{fig:fig1} (b) illustrated within a dashed box.

The data collected from syndrome measurement over several cycles is processed by a classical syndrome decoding method. This prescribes adequate corrections to fix physical errors in data qubits and restore the logical state of the lattice.
The construction of an efficient and scalable syndrome decoder is a challenging computational problem and has recently been the focus of intensive research \cite{skoric_parallel_2023, battistel_real-time_2023}. 
One of the leading syndrome decoder algorithms, MWPM, calculates corrections by matching pairs of changed stabilisers.
It has received extensive development in many square lattice surface code studies \cite{fowler_surface_2012, fowler_minimum_2015, acharya_suppressing_2023, fowler_towards_2012, wang_threshold_2010, higgott_pymatching_2022}. 
Chamberland \textit{et al.} implemented the MWPM algorithm to the original HH layout to compute logical error rate curves for both $X$ and $Z$ logical errors \cite{chamberland_topological_2020}. 

\begin{figure*}[t]
    \centering
    \includegraphics[width=\linewidth]{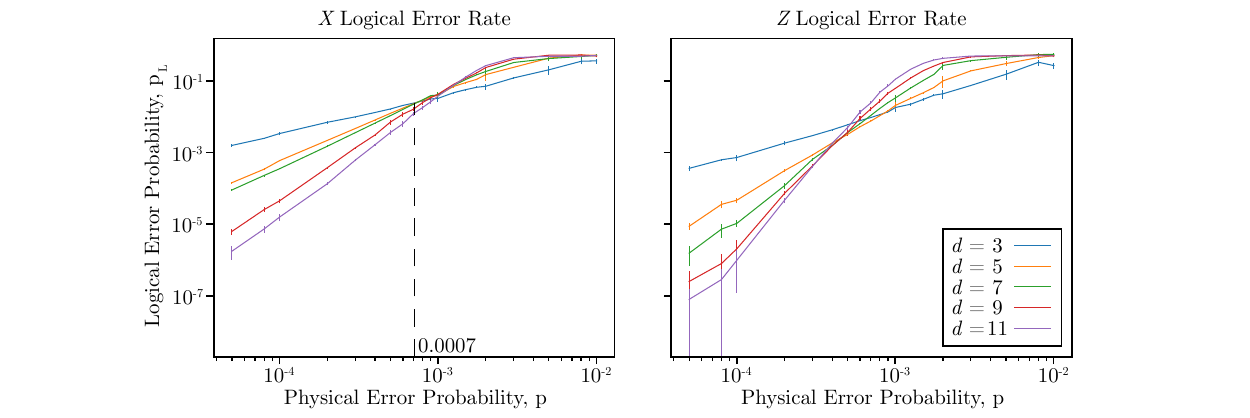}
    \caption{\textbf{Benchmarking of the adjusted Heavy Hexagon Code with MWPM.} Both the threshold and psuedo-threshold for $X$ logical errors (left) and $Z$ logical errors (right) for the adjusted HH code are shown; decoded by MWPM as implemeted by PyMatching. Error bars are assigned with a probit corresponding to $97.5\%$.}
    \label{fig:IBMHHSim}
\end{figure*}

We benchmarked the adjusted HH code using the MWPM decoder from the Python package PyMatching \cite{higgott_pymatching_2022} and compared it to the Chamberland \textit{et al.} work. 
In Figure \ref{fig:IBMHHSim}, odd distances, $d$, of the code between three to eleven are tested and the lowest clear crossover point can be seen at approximately 0.0007 in the $X$ logical error plot on the left. 
This will be the benchmark for thresholds for the adjusted HH QEC code, as only distances three and five are experimentally tested.
Note that the threshold if computed based on increaseing code distance would be slightly higher (\~0.001).
We used the MWPM as implemented by PyMatching to confirm Chamberland \textit{et al.}'s $X$ logical error threshold of 0.0045, and a very similar threshold of 0.005 was found. Details of this can be found in the Supplementary Material Section \ref{sec:MWPM}.
The addition of $2d-2$ extra ancilla qubits and $2d-2$ of CNOTs has lowered the threshold physical error probability further by a small amount.

Despite promising performance, it has been regularly discussed that the MWPM algorithm may not be fast enough for quantum state coherence times on current devices \cite{delfosse_how_2023, dennis_topological_2002, torlai_neural_2017, chamberland_deep_2018}. 
Even the best adaptations of this algorithm are slow in the large distance regime of QEC codes. 
The development of fast and scalable syndrome decoders is beginning its investigation, with recent proposals attempting to address the real-time decoding challenge \cite{battistel_real-time_2023}. 
Machine Learning (ML) based syndrome decoder construction has gained significant momentum in recent years, with some studies indicating that a faster and scalable syndrome decoding method may be possible by leveraging the computational efficiency and flexibility of ANN algorithms.  
For example, in the case of square surface code lattices, it has been shown that ANN syndrome decoders can offer highly promising performance when suggesting suitable corrections \cite{torlai_neural_2017, krastanov_deep_2017, varsamopoulos_decoding_2017, varsamopoulos_comparing_2020, baireuther_machine-learning-assisted_2018, davaasuren_general_2020}, including testing on experimental data \cite{bausch_learning_2023}.
The low-level decoders developed in \cite{torlai_neural_2017, krastanov_deep_2017, varsamopoulos_decoding_2017} were built in a similar manner to this work. They each show the ability of an ANN to learn the relationship between syndrome data and corrections after being given multiple training instances.
Many ANN varieties have been developed for square surface codes, including dense, Feed Forward Neural Networks (FFNN), Long Short Term Memory Networks (LSTM), and Convolutional Neural Networks (CNN). 
Varsamopolous \textit{et al.} showed that although slower than the FFNN, the LSTM was more accurate at decoding on average and both were faster and more accurate than the MWPM baseline \cite{varsamopoulos_comparing_2020}. 
Gicev \textit{et al.} and Meinerz \textit{et al.} have independently shown that implementing Convolutional layers allows an ANN decoder to be compatible with larger code distances unseen in training \cite{gicev_scalable_2023, meinerz_scalable_2022}. 
These results showed that an ANN syndrome decoder is able to fit any size QEC square surface code. 
In terms of the HH architecture, there is no current experimental implementation on IBM devices with the required adjustment of the code boundary described in Supplementary Material Section \ref{sec:adjustment}. 
Recent theoretical studies explored the implementation of dense ANN and CNN based decoders for the original HH code proposed by Chamberland \textit{et al.}, however their work is not directly applicable to the IBM hardware \cite{bhoumik_efficient_2022, li_convolutional-neural-network-based_2023}. 
Our work is the first to implement and benchmark an ANN decoder on the adjusted HH code through theoretical simulations and experimental measurements.

\begin{figure*}[t]
    \centering
    \includegraphics[width=\linewidth]{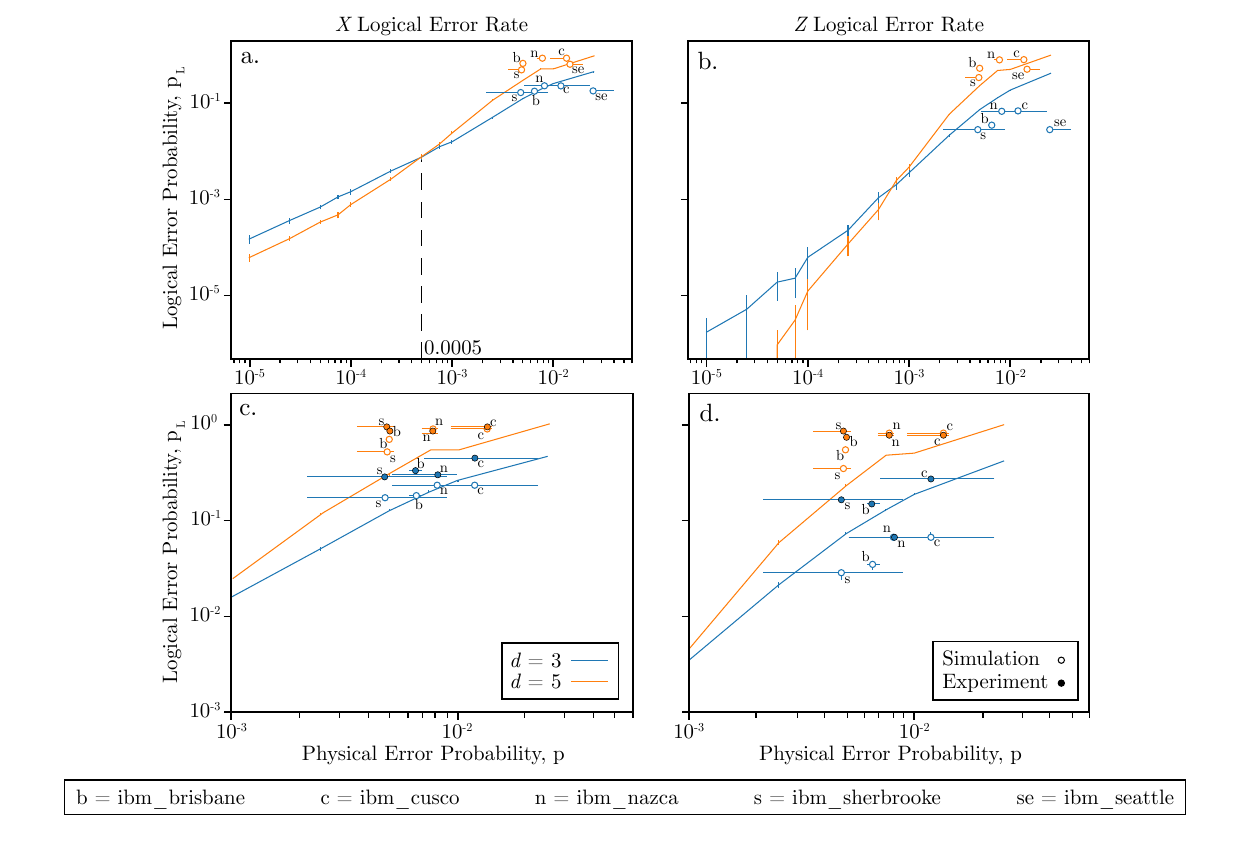}
    \caption{\textbf{Neural Network Decoder Implementation on Adjusted Heavy Hexagon Code.} Threshold plot for the adjusted HH code decoded by an ANN showing error rates of the $X$ logical operator (a.) and $Z$ logical operator (b.). Each point refers to an error model derived for each IBM device. The horizontal value of the points shown are the overall error rate of the specific sub-graph location chosen, and the horizontal uncertainty shows the range of overall error rates of each possible sub-graph location on each device, with the point placed on the median heuristic sub-graph score. Vertical confidence is found with a probit corresponding to $95.0\%$. c. and d. refer to the HH QEC code experimental circuit plots. The top right hand corner of a. and b. is enlarged, and the points which refer to the circuits running on the IBM devices are also marked. Unfilled circles refer to the simulated noise model corrections and filled circles refer to the transpiled circuits run on devices. }
    \label{fig:IBMHHNN}
\end{figure*}

The ANN decoder developed in this work was constructed using the Python package TensorFlow \cite{abadi_tensorflow_2016}.  
The decoder consists of an input, two hidden, and an output layer. 
The choice for the number of hidden layers is based on the decoder performance. Limited overfitting of training data occurred when two hidden layers were included. 
Utilising exclusively dense layers is the simplest layer structure of a neural network and requires no additional pruning or alterations \cite{varsamopoulos_decoding_2017}. 
This methodology allows for the quick proof of concept construction of an ANN syndrome decoder for physical devices, and can give suitable corrections without much pre-/post-processing. 
Given that the input layer takes the entirety of the syndrome measurement at once, there is no need to explicitly distinguish between bulk stabilisers and boundary stabilisers when training the network.
The network is able to learn the direct relationship between observed syndrome patterns and appropriate corrections without needing to perform auxiliary tasks after corrections are applied --- similar to the MWPM algorithm.
The MWPM algorithm can provide exact corrections by pairing $-1$ eigenvalue stabilisers without the need for any pre-/post-processing, but is lacking in the speed of suggestion, especially as the distance of the code increases. 

The ANNs developed for this work are built with dense layers, meaning each neuron within each layer is intricately connected with each neuron in the previous layers.
In Figure \ref{fig:fig1} (e.) a dense ANN is illustrated, where the number of neurons in each layer linearly decreases from the input layer to the output layer. 
The size of the input layer is adjusted to feed in each $X$ and $Z$ stabiliser measurement separately, for each measurement cycle. The size of the output layer allows a value for both $X$ and $Z$ errors for each physical qubit. 
At the smallest distance, three, the size of input and output layers are the same, however the input layer size grows significantly faster than the output layer when the distance of the code is increased.
Each entry in the input corresponds to a single stabiliser measurement, with the total equalling the number of stabilisers, $n$ multiplied by the number of cycles, $d$; $\frac{d}{2} \left(d^2 + 2d - 3\right)$. Similarly, each output pair corresponds to a single data qubit requiring $X$ and $Z$ correction respectively. The total output size is $2d^2$.

Each layer is activated with the ReLU activation function, excluding the final layer which incorporates a Sigmoid function, to return values between $0$ and $1$. 
The BinaryCrossEntropy loss function was used, allowing the network output to be interpreted as a probability that an error was present at each qubit.
Each value in the output of the final layer will be a value between $0$ and $1$, which are then processed in two ways. 
First, the values are truncated, such that each value in the correction suggestion is exactly $0$ or $1$, which corresponds to a given correction being not required or required, respectively. 
If this correction is consistent with the final syndrome measurement cycle, the truncated prediction is kept. 
If not, the predictions given are sampled using a Bernoulli Trial, and this is repeated until an appropriate correction is given \cite{krastanov_deep_2017}.
Sampling of a prediction could take many re-tries if the network is uncertain with its prediction. 
Therefore a cut-off point is used, where after $n$ re-samples, if no appropriate correction is given, re-sampling is stopped and it is assumed that a logical error has occurred in that instance \cite{krastanov_deep_2017}. Although there is theoretically a 50\% chance that a logical error has occurred, for benchmarking purposes, the occurrence of a logical error is assumed and the additional logical errors are reflected in Figure \ref{fig:IBMHHNN}.
Re-sampling can be a major overhead computationally, furthering the need to cut-off early, before qubits decohere within the structure. 
Given that this work only considered small distances of the HH QEC code, truncation of predictions often produced an appropriate correction --- not always requiring re-sampling. 
The coherence time of current qubits is on the order of microseconds and this work's re-sample time is also on the order of microseconds, forcing re-sampling to be avoided as much as possible \cite{ibm_ibm_2022}.
This dense ANN methodology is fast enough to produce corrections within the coherence time of physical qubits in the lattice for these small distances \cite{ibm_ibm_2022}.
The time taken to find corrections increases with code distance, and may not be appropriate for large distance codes.
Instead, CNN techniques can be employed for decoding large distance codes \cite{meinerz_scalable_2022, gicev_scalable_2023, chamberland_techniques_2023, ueno_neo-qec_2022}.

Our ANN decoder was rigorously trained on millions of simulated noise patterns, using uniform depolarising Pauli channels. 
The uniform depolarising noise model was simulated with an even chance, $\frac{p}{3}$, to select from the three Pauli gate errors $X$, $Y$ and $Z$.
Each qubit can experience each of these errors, and each CNOT on the lattice can experience some tensor product of two Pauli gate errors and the identity, excluding $I \otimes I$. 
No bias or other error factors were included in this training. 
During training, circuits were modelled such that when Pauli errors occur on a state, $\ket{\psi}$, it may be denoted as $E \ket{\psi}$ where $E$ is the combination of errors on a single qubit. 
The goal of error correction is to detect and apply the appropriate correction to $\ket{\psi}$ to turn the string of errors $E$ into the identity, $I$, or to return the lattice to an equivalent logical state. 
We compute a correction $E_c$ such that the correction succeeds if $E_c E \in G$ where $G$ is the corresponding gauge group.
This is simulated within this work by tracking each error which occurs on every qubit and multiplying the Pauli gate errors, where two of the same give the identity; $X^2 = Y^2 = Z^2 = I$, and $XZ = ZX = Y$ up to a global phase.

The ANN decoder developed in our work provides appropriate corrections based on the syndrome measurements over $d$ cycles of the adjusted HH code. 
The ANN is able to functionally learn how stabiliser inversions are related to error chains within the lattice, including on the boundaries of the lattice where chains abruptly end. 
This work has explicitly shown that a dense ANN syndrome decoder can input an exact stabiliser syndrome measurement and return a prediction related to an appropriate correction on par or better than suggestions from the MWPM algorithm. 

First evaluation of the model was done with a similar error model to the training; an uniform depolarising noise model.
The underlying physical error, $p$ was varied to test the performance at different rates. 
Further, the decoders were then tested on imported device error models from IBM quantum experience; each physical error rate was given for qubits and two-qubit gates which was then used as the underlying error probability $p$. 
This was implemented for each individual qubit and CNOT, instead of uniformly across the lattice. 
Finally, the circuits as defined in Figure \ref{fig:fig1} (d) were constructed to fit distance three and five HH QEC codes and executed on multiple IBM devices. 
Figure \ref{fig:IBMHHNN} (a) and (b) display experimental and theoretical results from our ANN syndrome decoder. The decoder is tested on a simulated lattice of qubits in the form of IBM devices which suffer from uniform circuit-based depolarising noise (blue and orange line plots) and also on device noise models derived from error rates provided by five of IBM quantum processors (marked open circle points). 
In these plots, similar crossover behaviour is observed, and thus it can be inferred that the ANN syndrome decoder is able to decode the HH QEC code with the same overall properties as the MWPM algorithm. 
Note that the threshold for the ANN syndrome decoder is approximately 0.0005 for $X$ logical errors, and hence reduced by a a small amount compared to the MWPM threshold of 0.0007 from Figure \ref{fig:IBMHHSim} (a). 
In future, more sophisticated ML-based syndrome decoders, such as CNN decoders, can be designed to improve the threshold and scale to larger distances \cite{davaasuren_general_2020, chamberland_techniques_2023, gicev_scalable_2023, li_convolutional-neural-network-based_2023}.

In Figures \ref{fig:IBMHHNN} (a) and (b), the blue circle markings correspond to distance three sub-graphs and orange for distance five. Each data point has an alphabetical label showing the name of IBM device: $b=ibm\_brisbane$, $c=ibm\_cusco$, $n=ibm\_nazca$, $s=ibm\_sherbrooke$, and $se=ibm\_seattle$. 
The horizontal uncertainty for each marking corresponds to the possible values of average physical error for each available sub-graph location, chosen with a heuristic described in Supplementary Material Section \ref{sec:sub-graph}, with the marking corresponding to the median location.
Interestingly, the markings are in the approximate region of the simulated noise curves. 
This suggests that the ANN syndrome decoder is likely to be able to decode actual noise approximately as well as simulated noise. 
Due to the above threshold error rates of current physical machines, distances above five were not tested, since this would only increase the logical error rate and may not provide additional insight.

Figure \ref{fig:IBMHHNN} (c) and (d) plot results based on direct measurements from the IBM quantum processors.
The plots show both device noise simulations (open circles) from Figure \ref{fig:IBMHHNN} (a) and (b), as well as experimental points (coloured circles) for a direct comparison. 
For the experimental points, the adjusted HH QEC code syndrome measurement circuits were created and run on physically realised IBM devices. 
Each circuit was initialised twice, once for $X$ measurements and once for $Z$ measurements, and 10,000 shots were run for each case.
The number of logical errors which occurred after the pass through of the ANN syndrome decoder was lower on average that the simulated noise models of the same devices for distance three, and roughly similar for distance five.
Given that the points are still all within the same area or lower, it would follow that if the devices error rates were below the threshold of approximately 0.0005, that increasing the distance of the code, and using a suitable ANN syndrome decoder, would facilitate fault-tolerant quantum computation \cite{acharya_suppressing_2023}.

Note that in Figure \ref{fig:IBMHHNN} (b), that the device derived error models seem to consistently provide lower logical error rates than equivalent uniform error models.
This suggests that there is some intricate phenomenon occurring which may be related to sub-graph location choice.
Compared to what is expected under the uniform noise model, this results in the reduction of the rate of $Z$ logical errors, which corrupt $X$ logical operator values.
This is not observed in Figure \ref{fig:IBMHHNN} (d) however, as the experimental data is not lower than the simulated uniform noise curve on average.
Crosstalk and relaxation errors are missing in the simulated noise model but are possibly present on the physically realised devices, perhaps leading to this variation between experiment and simulation \cite{tomita_low-distance_2014, chen_calibrated_2022}.

Despite the expeditious advances in quantum hardware, fault-tolerant quantum computation is still distant on the horizon. 
However, this work, for the first time, showed that the adjusted HH code which matches the IBM quantum machine structure, is able to be decoded by both the MWPM algorithm, and an ANN syndrome decoder.
A dense ANN was shown to be compatible with the adjusted HH code and to perform experimentally in accordance to the error rates present on the devices. 
The experimental results in this work showed that stabiliser circuit decoding approximately followed the theoretical curve's trend. 
It is therefore likely that lowering the physical error rate below the threshold will allow for arbitrary suppression of logical errors with code distance increase. 
This work's dense style ANN lays the foundation of ANN decoding on physically realised IBM quantum machines.

In future, our work could be extended with the benchmarking of larger distance code implementations on IBM devices to demonstrate the expected drop in logical error rates with respect to code distance. 
However, this would require larger physical devices and devices with error rates below the code threshold. 
A second line of study could be to implement and test more sophisticated ML-based decoders --- such as CNN syndrome decoders --- on quantum devices. 
In summary, our work has opened new avenues for experimentally realised, ML-based syndrome decoder implementation on quantum processors. 
This will be instrumental in realising fault-tolerant quantum computing in the near future, where larger size and lower error rate devices are anticipated to be available.\\

\textbf{Acknowledgements:}

The research was supported by the University of Melbourne through the establishment of the IBM Quantum Network Hub at the University. 
Computational resources were provided by the National Computing Infrastructure (NCI) and the Pawsey Supercomputing Research Center through the National Computational Merit Allocation Scheme (NCMAS).\\

\textbf{Author Contributions:}

M.U. planned and supervised the project. B.H. developed the Artificial Neural Network Decoders with input from S.G. B.H. carried out all simulations, benchmarking and experimental implementation. All authors contributed to the analysis of data. B.H. and M.U. wrote the manuscript with contributions from S.G.

\bibliography{references}

\begin{thebibliography}{50}%
\makeatletter
\providecommand \@ifxundefined [1]{%
 \@ifx{#1\undefined}
}%
\providecommand \@ifnum [1]{%
 \ifnum #1\expandafter \@firstoftwo
 \else \expandafter \@secondoftwo
 \fi
}%
\providecommand \@ifx [1]{%
 \ifx #1\expandafter \@firstoftwo
 \else \expandafter \@secondoftwo
 \fi
}%
\providecommand \natexlab [1]{#1}%
\providecommand \enquote  [1]{``#1''}%
\providecommand \bibnamefont  [1]{#1}%
\providecommand \bibfnamefont [1]{#1}%
\providecommand \citenamefont [1]{#1}%
\providecommand \href@noop [0]{\@secondoftwo}%
\providecommand \href [0]{\begingroup \@sanitize@url \@href}%
\providecommand \@href[1]{\@@startlink{#1}\@@href}%
\providecommand \@@href[1]{\endgroup#1\@@endlink}%
\providecommand \@sanitize@url [0]{\catcode `\\12\catcode `\$12\catcode `\&12\catcode `\#12\catcode `\^12\catcode `\_12\catcode `\%12\relax}%
\providecommand \@@startlink[1]{}%
\providecommand \@@endlink[0]{}%
\providecommand \url  [0]{\begingroup\@sanitize@url \@url }%
\providecommand \@url [1]{\endgroup\@href {#1}{\urlprefix }}%
\providecommand \urlprefix  [0]{URL }%
\providecommand \Eprint [0]{\href }%
\providecommand \doibase [0]{https://doi.org/}%
\providecommand \selectlanguage [0]{\@gobble}%
\providecommand \bibinfo  [0]{\@secondoftwo}%
\providecommand \bibfield  [0]{\@secondoftwo}%
\providecommand \translation [1]{[#1]}%
\providecommand \BibitemOpen [0]{}%
\providecommand \bibitemStop [0]{}%
\providecommand \bibitemNoStop [0]{.\EOS\space}%
\providecommand \EOS [0]{\spacefactor3000\relax}%
\providecommand \BibitemShut  [1]{\csname bibitem#1\endcsname}%
\let\auto@bib@innerbib\@empty
\bibitem [{\citenamefont {Collins}\ and\ \citenamefont {Nay}(2022)}]{collins_ibm_2022}%
  \BibitemOpen
  \bibfield  {author} {\bibinfo {author} {\bibfnamefont {H.}~\bibnamefont {Collins}}\ and\ \bibinfo {author} {\bibfnamefont {C.}~\bibnamefont {Nay}},\ }\href {https://newsroom.ibm.com/2022-11-09-IBM-Unveils-400-Qubit-Plus-Quantum-Processor-and-Next-Generation-IBM-Quantum-System-Two} {\bibinfo {title} {{IBM} {Unveils} 400 {Qubit}-{Plus} {Quantum} {Processor} and {Next}-{Generation} {IBM} {Quantum} {System} {Two}}} (\bibinfo {year} {2022})\BibitemShut {NoStop}%
\bibitem [{\citenamefont {Madsen}\ \emph {et~al.}(2022)\citenamefont {Madsen}, \citenamefont {Laudenbach}, \citenamefont {Askarani}, \citenamefont {Rortais}, \citenamefont {Vincent}, \citenamefont {Bulmer}, \citenamefont {Miatto}, \citenamefont {Neuhaus}, \citenamefont {Helt}, \citenamefont {Collins}, \citenamefont {Lita}, \citenamefont {Gerrits}, \citenamefont {Nam}, \citenamefont {Vaidya}, \citenamefont {Menotti}, \citenamefont {Dhand}, \citenamefont {Vernon}, \citenamefont {Quesada},\ and\ \citenamefont {Lavoie}}]{madsen_quantum_2022}%
  \BibitemOpen
  \bibfield  {author} {\bibinfo {author} {\bibfnamefont {L.~S.}\ \bibnamefont {Madsen}}, \bibinfo {author} {\bibfnamefont {F.}~\bibnamefont {Laudenbach}}, \bibinfo {author} {\bibfnamefont {M.~F.}\ \bibnamefont {Askarani}}, \bibinfo {author} {\bibfnamefont {F.}~\bibnamefont {Rortais}}, \bibinfo {author} {\bibfnamefont {T.}~\bibnamefont {Vincent}}, \bibinfo {author} {\bibfnamefont {J.~F.~F.}\ \bibnamefont {Bulmer}}, \bibinfo {author} {\bibfnamefont {F.~M.}\ \bibnamefont {Miatto}}, \bibinfo {author} {\bibfnamefont {L.}~\bibnamefont {Neuhaus}}, \bibinfo {author} {\bibfnamefont {L.~G.}\ \bibnamefont {Helt}}, \bibinfo {author} {\bibfnamefont {M.~J.}\ \bibnamefont {Collins}}, \bibinfo {author} {\bibfnamefont {A.~E.}\ \bibnamefont {Lita}}, \bibinfo {author} {\bibfnamefont {T.}~\bibnamefont {Gerrits}}, \bibinfo {author} {\bibfnamefont {S.~W.}\ \bibnamefont {Nam}}, \bibinfo {author} {\bibfnamefont {V.~D.}\ \bibnamefont {Vaidya}}, \bibinfo {author} {\bibfnamefont {M.}~\bibnamefont {Menotti}}, \bibinfo {author}
  {\bibfnamefont {I.}~\bibnamefont {Dhand}}, \bibinfo {author} {\bibfnamefont {Z.}~\bibnamefont {Vernon}}, \bibinfo {author} {\bibfnamefont {N.}~\bibnamefont {Quesada}},\ and\ \bibinfo {author} {\bibfnamefont {J.}~\bibnamefont {Lavoie}},\ }\href {https://doi.org/10.1038/s41586-022-04725-x} {\bibfield  {journal} {\bibinfo  {journal} {Nature}\ }\textbf {\bibinfo {volume} {606}},\ \bibinfo {pages} {75} (\bibinfo {year} {2022})}\BibitemShut {NoStop}%
\bibitem [{\citenamefont {Barnes}\ \emph {et~al.}(2022)\citenamefont {Barnes}, \citenamefont {Battaglino}, \citenamefont {Bloom}, \citenamefont {Cassella}, \citenamefont {Coxe}, \citenamefont {Crisosto}, \citenamefont {King}, \citenamefont {Kondov}, \citenamefont {Kotru}, \citenamefont {Larsen}, \citenamefont {Lauigan}, \citenamefont {Lester}, \citenamefont {McDonald}, \citenamefont {Megidish}, \citenamefont {Narayanaswami}, \citenamefont {Nishiguchi}, \citenamefont {Notermans}, \citenamefont {Peng}, \citenamefont {Ryou}, \citenamefont {Wu},\ and\ \citenamefont {Yarwood}}]{barnes_assembly_2022}%
  \BibitemOpen
  \bibfield  {author} {\bibinfo {author} {\bibfnamefont {K.}~\bibnamefont {Barnes}}, \bibinfo {author} {\bibfnamefont {P.}~\bibnamefont {Battaglino}}, \bibinfo {author} {\bibfnamefont {B.~J.}\ \bibnamefont {Bloom}}, \bibinfo {author} {\bibfnamefont {K.}~\bibnamefont {Cassella}}, \bibinfo {author} {\bibfnamefont {R.}~\bibnamefont {Coxe}}, \bibinfo {author} {\bibfnamefont {N.}~\bibnamefont {Crisosto}}, \bibinfo {author} {\bibfnamefont {J.~P.}\ \bibnamefont {King}}, \bibinfo {author} {\bibfnamefont {S.~S.}\ \bibnamefont {Kondov}}, \bibinfo {author} {\bibfnamefont {K.}~\bibnamefont {Kotru}}, \bibinfo {author} {\bibfnamefont {S.~C.}\ \bibnamefont {Larsen}}, \bibinfo {author} {\bibfnamefont {J.}~\bibnamefont {Lauigan}}, \bibinfo {author} {\bibfnamefont {B.~J.}\ \bibnamefont {Lester}}, \bibinfo {author} {\bibfnamefont {M.}~\bibnamefont {McDonald}}, \bibinfo {author} {\bibfnamefont {E.}~\bibnamefont {Megidish}}, \bibinfo {author} {\bibfnamefont {S.}~\bibnamefont {Narayanaswami}}, \bibinfo {author} {\bibfnamefont
  {C.}~\bibnamefont {Nishiguchi}}, \bibinfo {author} {\bibfnamefont {R.}~\bibnamefont {Notermans}}, \bibinfo {author} {\bibfnamefont {L.~S.}\ \bibnamefont {Peng}}, \bibinfo {author} {\bibfnamefont {A.}~\bibnamefont {Ryou}}, \bibinfo {author} {\bibfnamefont {T.-Y.}\ \bibnamefont {Wu}},\ and\ \bibinfo {author} {\bibfnamefont {M.}~\bibnamefont {Yarwood}},\ }\href {https://doi.org/10.1038/s41467-022-29977-z} {\bibfield  {journal} {\bibinfo  {journal} {Nature Communications}\ }\textbf {\bibinfo {volume} {13}},\ \bibinfo {pages} {2779} (\bibinfo {year} {2022})}\BibitemShut {NoStop}%
\bibitem [{\citenamefont {Beverland}\ \emph {et~al.}(2022)\citenamefont {Beverland}, \citenamefont {Murali}, \citenamefont {Troyer}, \citenamefont {Svore}, \citenamefont {Hoefler}, \citenamefont {Kliuchnikov}, \citenamefont {Low}, \citenamefont {Soeken}, \citenamefont {Sundaram},\ and\ \citenamefont {Vaschillo}}]{beverland_assessing_2022}%
  \BibitemOpen
  \bibfield  {author} {\bibinfo {author} {\bibfnamefont {M.~E.}\ \bibnamefont {Beverland}}, \bibinfo {author} {\bibfnamefont {P.}~\bibnamefont {Murali}}, \bibinfo {author} {\bibfnamefont {M.}~\bibnamefont {Troyer}}, \bibinfo {author} {\bibfnamefont {K.~M.}\ \bibnamefont {Svore}}, \bibinfo {author} {\bibfnamefont {T.}~\bibnamefont {Hoefler}}, \bibinfo {author} {\bibfnamefont {V.}~\bibnamefont {Kliuchnikov}}, \bibinfo {author} {\bibfnamefont {G.~H.}\ \bibnamefont {Low}}, \bibinfo {author} {\bibfnamefont {M.}~\bibnamefont {Soeken}}, \bibinfo {author} {\bibfnamefont {A.}~\bibnamefont {Sundaram}},\ and\ \bibinfo {author} {\bibfnamefont {A.}~\bibnamefont {Vaschillo}},\ }\href {https://arxiv.org/abs/2211.07629v1} {\bibinfo {title} {Assessing requirements to scale to practical quantum advantage}} (\bibinfo {year} {2022})\BibitemShut {NoStop}%
\bibitem [{\citenamefont {Kim}\ \emph {et~al.}(2023)\citenamefont {Kim}, \citenamefont {Eddins}, \citenamefont {Anand}, \citenamefont {Wei}, \citenamefont {van~den Berg}, \citenamefont {Rosenblatt}, \citenamefont {Nayfeh}, \citenamefont {Wu}, \citenamefont {Zaletel}, \citenamefont {Temme},\ and\ \citenamefont {Kandala}}]{kim_evidence_2023}%
  \BibitemOpen
  \bibfield  {author} {\bibinfo {author} {\bibfnamefont {Y.}~\bibnamefont {Kim}}, \bibinfo {author} {\bibfnamefont {A.}~\bibnamefont {Eddins}}, \bibinfo {author} {\bibfnamefont {S.}~\bibnamefont {Anand}}, \bibinfo {author} {\bibfnamefont {K.~X.}\ \bibnamefont {Wei}}, \bibinfo {author} {\bibfnamefont {E.}~\bibnamefont {van~den Berg}}, \bibinfo {author} {\bibfnamefont {S.}~\bibnamefont {Rosenblatt}}, \bibinfo {author} {\bibfnamefont {H.}~\bibnamefont {Nayfeh}}, \bibinfo {author} {\bibfnamefont {Y.}~\bibnamefont {Wu}}, \bibinfo {author} {\bibfnamefont {M.}~\bibnamefont {Zaletel}}, \bibinfo {author} {\bibfnamefont {K.}~\bibnamefont {Temme}},\ and\ \bibinfo {author} {\bibfnamefont {A.}~\bibnamefont {Kandala}},\ }\href {https://doi.org/10.1038/s41586-023-06096-3} {\bibfield  {journal} {\bibinfo  {journal} {Nature}\ }\textbf {\bibinfo {volume} {618}},\ \bibinfo {pages} {500} (\bibinfo {year} {2023})}\BibitemShut {NoStop}%
\bibitem [{\citenamefont {Endo}\ \emph {et~al.}(2018)\citenamefont {Endo}, \citenamefont {Benjamin},\ and\ \citenamefont {Li}}]{endo_practical_2018}%
  \BibitemOpen
  \bibfield  {author} {\bibinfo {author} {\bibfnamefont {S.}~\bibnamefont {Endo}}, \bibinfo {author} {\bibfnamefont {S.~C.}\ \bibnamefont {Benjamin}},\ and\ \bibinfo {author} {\bibfnamefont {Y.}~\bibnamefont {Li}},\ }\href {https://doi.org/10.1103/PhysRevX.8.031027} {\bibfield  {journal} {\bibinfo  {journal} {Physical Review X}\ }\textbf {\bibinfo {volume} {8}},\ \bibinfo {pages} {031027} (\bibinfo {year} {2018})}\BibitemShut {NoStop}%
\bibitem [{\citenamefont {Strikis}\ \emph {et~al.}(2021)\citenamefont {Strikis}, \citenamefont {Qin}, \citenamefont {Chen}, \citenamefont {Benjamin},\ and\ \citenamefont {Li}}]{strikis_learning-based_2021}%
  \BibitemOpen
  \bibfield  {author} {\bibinfo {author} {\bibfnamefont {A.}~\bibnamefont {Strikis}}, \bibinfo {author} {\bibfnamefont {D.}~\bibnamefont {Qin}}, \bibinfo {author} {\bibfnamefont {Y.}~\bibnamefont {Chen}}, \bibinfo {author} {\bibfnamefont {S.~C.}\ \bibnamefont {Benjamin}},\ and\ \bibinfo {author} {\bibfnamefont {Y.}~\bibnamefont {Li}},\ }\href {https://doi.org/10.1103/PRXQuantum.2.040330} {\bibfield  {journal} {\bibinfo  {journal} {PRX Quantum}\ }\textbf {\bibinfo {volume} {2}},\ \bibinfo {pages} {040330} (\bibinfo {year} {2021})}\BibitemShut {NoStop}%
\bibitem [{\citenamefont {Bravyi}\ \emph {et~al.}(2021)\citenamefont {Bravyi}, \citenamefont {Sheldon}, \citenamefont {Kandala}, \citenamefont {Mckay},\ and\ \citenamefont {Gambetta}}]{bravyi_mitigating_2021}%
  \BibitemOpen
  \bibfield  {author} {\bibinfo {author} {\bibfnamefont {S.}~\bibnamefont {Bravyi}}, \bibinfo {author} {\bibfnamefont {S.}~\bibnamefont {Sheldon}}, \bibinfo {author} {\bibfnamefont {A.}~\bibnamefont {Kandala}}, \bibinfo {author} {\bibfnamefont {D.~C.}\ \bibnamefont {Mckay}},\ and\ \bibinfo {author} {\bibfnamefont {J.~M.}\ \bibnamefont {Gambetta}},\ }\href {https://doi.org/10.1103/PhysRevA.103.042605} {\bibfield  {journal} {\bibinfo  {journal} {Physical Review A}\ }\textbf {\bibinfo {volume} {103}},\ \bibinfo {pages} {042605} (\bibinfo {year} {2021})}\BibitemShut {NoStop}%
\bibitem [{\citenamefont {Shor}(1995)}]{shor_scheme_1995}%
  \BibitemOpen
  \bibfield  {author} {\bibinfo {author} {\bibfnamefont {P.~W.}\ \bibnamefont {Shor}},\ }\href {https://doi.org/10.1103/PhysRevA.52.R2493} {\bibfield  {journal} {\bibinfo  {journal} {Physical Review A}\ }\textbf {\bibinfo {volume} {52}},\ \bibinfo {pages} {R2493} (\bibinfo {year} {1995})}\BibitemShut {NoStop}%
\bibitem [{\citenamefont {Kitaev}(2003)}]{kitaev_fault-tolerant_2003}%
  \BibitemOpen
  \bibfield  {author} {\bibinfo {author} {\bibfnamefont {A.~Y.}\ \bibnamefont {Kitaev}},\ }\href {https://doi.org/10.1016/S0003-4916(02)00018-0} {\bibfield  {journal} {\bibinfo  {journal} {Annals of Physics}\ }\textbf {\bibinfo {volume} {303}},\ \bibinfo {pages} {2} (\bibinfo {year} {2003})}\BibitemShut {NoStop}%
\bibitem [{\citenamefont {Aharonov}\ and\ \citenamefont {Ben-Or}(2008)}]{aharonov_fault-tolerant_2008}%
  \BibitemOpen
  \bibfield  {author} {\bibinfo {author} {\bibfnamefont {D.}~\bibnamefont {Aharonov}}\ and\ \bibinfo {author} {\bibfnamefont {M.}~\bibnamefont {Ben-Or}},\ }\href {https://doi.org/10.1137/S0097539799359385} {\bibfield  {journal} {\bibinfo  {journal} {SIAM Journal on Computing}\ }\textbf {\bibinfo {volume} {38}},\ \bibinfo {pages} {1207} (\bibinfo {year} {2008})}\BibitemShut {NoStop}%
\bibitem [{\citenamefont {Knill}\ \emph {et~al.}(1998)\citenamefont {Knill}, \citenamefont {Laflamme},\ and\ \citenamefont {Zurek}}]{knill_resilient_1998}%
  \BibitemOpen
  \bibfield  {author} {\bibinfo {author} {\bibfnamefont {E.}~\bibnamefont {Knill}}, \bibinfo {author} {\bibfnamefont {R.}~\bibnamefont {Laflamme}},\ and\ \bibinfo {author} {\bibfnamefont {W.~H.}\ \bibnamefont {Zurek}},\ }\href {https://doi.org/10.1098/rspa.1998.0166} {\bibfield  {journal} {\bibinfo  {journal} {Proceedings of the Royal Society of London. Series A: Mathematical, Physical and Engineering Sciences}\ }\textbf {\bibinfo {volume} {454}},\ \bibinfo {pages} {365} (\bibinfo {year} {1998})}\BibitemShut {NoStop}%
\bibitem [{\citenamefont {Gottesman}(1997)}]{gottesman_stabilizer_1997}%
  \BibitemOpen
  \bibfield  {author} {\bibinfo {author} {\bibfnamefont {D.}~\bibnamefont {Gottesman}},\ }\href {https://arxiv.org/abs/quant-ph/9705052v1} {\bibinfo {title} {Stabilizer {Codes} and {Quantum} {Error} {Correction}}} (\bibinfo {year} {1997})\BibitemShut {NoStop}%
\bibitem [{\citenamefont {Steane}(1997)}]{steane_active_1997}%
  \BibitemOpen
  \bibfield  {author} {\bibinfo {author} {\bibfnamefont {A.~M.}\ \bibnamefont {Steane}},\ }\href {https://doi.org/10.1103/PhysRevLett.78.2252} {\bibfield  {journal} {\bibinfo  {journal} {Physical Review Letters}\ }\textbf {\bibinfo {volume} {78}},\ \bibinfo {pages} {2252} (\bibinfo {year} {1997})}\BibitemShut {NoStop}%
\bibitem [{\citenamefont {Kitaev}(1997)}]{kitaev_quantum_1997}%
  \BibitemOpen
  \bibfield  {author} {\bibinfo {author} {\bibfnamefont {A.~Y.}\ \bibnamefont {Kitaev}},\ }in\ \href {https://doi.org/10.1007/978-1-4615-5923-8_19} {\emph {\bibinfo {booktitle} {Quantum {Communication}, {Computing}, and {Measurement}}}},\ \bibinfo {editor} {edited by\ \bibinfo {editor} {\bibfnamefont {O.}~\bibnamefont {Hirota}}, \bibinfo {editor} {\bibfnamefont {A.~S.}\ \bibnamefont {Holevo}},\ and\ \bibinfo {editor} {\bibfnamefont {C.~M.}\ \bibnamefont {Caves}}}\ (\bibinfo  {publisher} {Springer US},\ \bibinfo {address} {Boston, MA},\ \bibinfo {year} {1997})\ pp.\ \bibinfo {pages} {181--188}\BibitemShut {NoStop}%
\bibitem [{\citenamefont {Fowler}\ \emph {et~al.}(2012{\natexlab{a}})\citenamefont {Fowler}, \citenamefont {Mariantoni}, \citenamefont {Martinis},\ and\ \citenamefont {Cleland}}]{fowler_surface_2012}%
  \BibitemOpen
  \bibfield  {author} {\bibinfo {author} {\bibfnamefont {A.~G.}\ \bibnamefont {Fowler}}, \bibinfo {author} {\bibfnamefont {M.}~\bibnamefont {Mariantoni}}, \bibinfo {author} {\bibfnamefont {J.~M.}\ \bibnamefont {Martinis}},\ and\ \bibinfo {author} {\bibfnamefont {A.~N.}\ \bibnamefont {Cleland}},\ }\href {https://doi.org/10.1103/PhysRevA.86.032324} {\bibfield  {journal} {\bibinfo  {journal} {Physical Review A}\ }\textbf {\bibinfo {volume} {86}},\ \bibinfo {pages} {032324} (\bibinfo {year} {2012}{\natexlab{a}})}\BibitemShut {NoStop}%
\bibitem [{\citenamefont {Varsamopoulos}\ \emph {et~al.}(2017)\citenamefont {Varsamopoulos}, \citenamefont {Criger},\ and\ \citenamefont {Bertels}}]{varsamopoulos_decoding_2017}%
  \BibitemOpen
  \bibfield  {author} {\bibinfo {author} {\bibfnamefont {S.}~\bibnamefont {Varsamopoulos}}, \bibinfo {author} {\bibfnamefont {B.}~\bibnamefont {Criger}},\ and\ \bibinfo {author} {\bibfnamefont {K.}~\bibnamefont {Bertels}},\ }\href {https://doi.org/10.1088/2058-9565/aa955a} {\bibfield  {journal} {\bibinfo  {journal} {Quantum Science and Technology}\ }\textbf {\bibinfo {volume} {3}},\ \bibinfo {pages} {015004} (\bibinfo {year} {2017})}\BibitemShut {NoStop}%
\bibitem [{\citenamefont {Meinerz}\ \emph {et~al.}(2022)\citenamefont {Meinerz}, \citenamefont {Park},\ and\ \citenamefont {Trebst}}]{meinerz_scalable_2022}%
  \BibitemOpen
  \bibfield  {author} {\bibinfo {author} {\bibfnamefont {K.}~\bibnamefont {Meinerz}}, \bibinfo {author} {\bibfnamefont {C.-Y.}\ \bibnamefont {Park}},\ and\ \bibinfo {author} {\bibfnamefont {S.}~\bibnamefont {Trebst}},\ }\href {https://doi.org/10.1103/PhysRevLett.128.080505} {\bibfield  {journal} {\bibinfo  {journal} {Physical Review Letters}\ }\textbf {\bibinfo {volume} {128}},\ \bibinfo {pages} {080505} (\bibinfo {year} {2022})}\BibitemShut {NoStop}%
\bibitem [{\citenamefont {Varsamopoulos}\ \emph {et~al.}(2020)\citenamefont {Varsamopoulos}, \citenamefont {Bertels},\ and\ \citenamefont {Almudever}}]{varsamopoulos_comparing_2020}%
  \BibitemOpen
  \bibfield  {author} {\bibinfo {author} {\bibfnamefont {S.}~\bibnamefont {Varsamopoulos}}, \bibinfo {author} {\bibfnamefont {K.}~\bibnamefont {Bertels}},\ and\ \bibinfo {author} {\bibfnamefont {C.~G.}\ \bibnamefont {Almudever}},\ }\href {https://doi.org/10.1109/TC.2019.2948612} {\bibfield  {journal} {\bibinfo  {journal} {IEEE Transactions on Computers}\ }\textbf {\bibinfo {volume} {69}},\ \bibinfo {pages} {300} (\bibinfo {year} {2020})}\BibitemShut {NoStop}%
\bibitem [{\citenamefont {Gicev}\ \emph {et~al.}(2023)\citenamefont {Gicev}, \citenamefont {Hollenberg},\ and\ \citenamefont {Usman}}]{gicev_scalable_2023}%
  \BibitemOpen
  \bibfield  {author} {\bibinfo {author} {\bibfnamefont {S.}~\bibnamefont {Gicev}}, \bibinfo {author} {\bibfnamefont {L.~C.~L.}\ \bibnamefont {Hollenberg}},\ and\ \bibinfo {author} {\bibfnamefont {M.}~\bibnamefont {Usman}},\ }\href {https://doi.org/10.22331/q-2023-07-12-1058} {\bibfield  {journal} {\bibinfo  {journal} {Quantum}\ }\textbf {\bibinfo {volume} {7}},\ \bibinfo {pages} {1058} (\bibinfo {year} {2023})}\BibitemShut {NoStop}%
\bibitem [{\citenamefont {Overwater}\ \emph {et~al.}(2022)\citenamefont {Overwater}, \citenamefont {Babaie},\ and\ \citenamefont {Sebastiano}}]{overwater_neural-network_2022}%
  \BibitemOpen
  \bibfield  {author} {\bibinfo {author} {\bibfnamefont {R.~W.~J.}\ \bibnamefont {Overwater}}, \bibinfo {author} {\bibfnamefont {M.}~\bibnamefont {Babaie}},\ and\ \bibinfo {author} {\bibfnamefont {F.}~\bibnamefont {Sebastiano}},\ }\href {https://doi.org/10.1109/TQE.2022.3174017} {\bibfield  {journal} {\bibinfo  {journal} {IEEE Transactions on Quantum Engineering}\ }\textbf {\bibinfo {volume} {3}},\ \bibinfo {pages} {1} (\bibinfo {year} {2022})}\BibitemShut {NoStop}%
\bibitem [{\citenamefont {Wagner}\ \emph {et~al.}(2020)\citenamefont {Wagner}, \citenamefont {Kampermann},\ and\ \citenamefont {Bruß}}]{wagner_symmetries_2020}%
  \BibitemOpen
  \bibfield  {author} {\bibinfo {author} {\bibfnamefont {T.}~\bibnamefont {Wagner}}, \bibinfo {author} {\bibfnamefont {H.}~\bibnamefont {Kampermann}},\ and\ \bibinfo {author} {\bibfnamefont {D.}~\bibnamefont {Bruß}},\ }\href {https://doi.org/10.1103/PhysRevA.102.042411} {\bibfield  {journal} {\bibinfo  {journal} {Physical Review A}\ }\textbf {\bibinfo {volume} {102}},\ \bibinfo {pages} {042411} (\bibinfo {year} {2020})}\BibitemShut {NoStop}%
\bibitem [{\citenamefont {Ni}(2020)}]{ni_neural_2020}%
  \BibitemOpen
  \bibfield  {author} {\bibinfo {author} {\bibfnamefont {X.}~\bibnamefont {Ni}},\ }\href {https://doi.org/10.22331/q-2020-08-24-310} {\bibfield  {journal} {\bibinfo  {journal} {Quantum}\ }\textbf {\bibinfo {volume} {4}},\ \bibinfo {pages} {310} (\bibinfo {year} {2020})}\BibitemShut {NoStop}%
\bibitem [{\citenamefont {Zhang}\ \emph {et~al.}(2023)\citenamefont {Zhang}, \citenamefont {Ren}, \citenamefont {Xi}, \citenamefont {Zhang}, \citenamefont {Yu}, \citenamefont {Liu}, \citenamefont {Zhang}, \citenamefont {Zhang},\ and\ \citenamefont {Zheng}}]{zhang_scalable_2023}%
  \BibitemOpen
  \bibfield  {author} {\bibinfo {author} {\bibfnamefont {M.}~\bibnamefont {Zhang}}, \bibinfo {author} {\bibfnamefont {X.}~\bibnamefont {Ren}}, \bibinfo {author} {\bibfnamefont {G.}~\bibnamefont {Xi}}, \bibinfo {author} {\bibfnamefont {Z.}~\bibnamefont {Zhang}}, \bibinfo {author} {\bibfnamefont {Q.}~\bibnamefont {Yu}}, \bibinfo {author} {\bibfnamefont {F.}~\bibnamefont {Liu}}, \bibinfo {author} {\bibfnamefont {H.}~\bibnamefont {Zhang}}, \bibinfo {author} {\bibfnamefont {S.}~\bibnamefont {Zhang}},\ and\ \bibinfo {author} {\bibfnamefont {Y.-C.}\ \bibnamefont {Zheng}},\ }\bibfield  {journal} {\bibinfo  {journal} {arXiv}\ }\href {https://doi.org/10.48550/arXiv.2305.15767} {10.48550/arXiv.2305.15767} (\bibinfo {year} {2023})\BibitemShut {NoStop}%
\bibitem [{\citenamefont {Bausch}\ \emph {et~al.}(2023)\citenamefont {Bausch}, \citenamefont {Senior}, \citenamefont {Heras}, \citenamefont {Edlich}, \citenamefont {Davies}, \citenamefont {Newman}, \citenamefont {Jones}, \citenamefont {Satzinger}, \citenamefont {Niu}, \citenamefont {Blackwell}, \citenamefont {Holland}, \citenamefont {Kafri}, \citenamefont {Atalaya}, \citenamefont {Gidney}, \citenamefont {Hassabis}, \citenamefont {Boixo}, \citenamefont {Neven},\ and\ \citenamefont {Kohli}}]{bausch_learning_2023}%
  \BibitemOpen
  \bibfield  {author} {\bibinfo {author} {\bibfnamefont {J.}~\bibnamefont {Bausch}}, \bibinfo {author} {\bibfnamefont {A.~W.}\ \bibnamefont {Senior}}, \bibinfo {author} {\bibfnamefont {F.~J.~H.}\ \bibnamefont {Heras}}, \bibinfo {author} {\bibfnamefont {T.}~\bibnamefont {Edlich}}, \bibinfo {author} {\bibfnamefont {A.}~\bibnamefont {Davies}}, \bibinfo {author} {\bibfnamefont {M.}~\bibnamefont {Newman}}, \bibinfo {author} {\bibfnamefont {C.}~\bibnamefont {Jones}}, \bibinfo {author} {\bibfnamefont {K.}~\bibnamefont {Satzinger}}, \bibinfo {author} {\bibfnamefont {M.~Y.}\ \bibnamefont {Niu}}, \bibinfo {author} {\bibfnamefont {S.}~\bibnamefont {Blackwell}}, \bibinfo {author} {\bibfnamefont {G.}~\bibnamefont {Holland}}, \bibinfo {author} {\bibfnamefont {D.}~\bibnamefont {Kafri}}, \bibinfo {author} {\bibfnamefont {J.}~\bibnamefont {Atalaya}}, \bibinfo {author} {\bibfnamefont {C.}~\bibnamefont {Gidney}}, \bibinfo {author} {\bibfnamefont {D.}~\bibnamefont {Hassabis}}, \bibinfo {author} {\bibfnamefont {S.}~\bibnamefont
  {Boixo}}, \bibinfo {author} {\bibfnamefont {H.}~\bibnamefont {Neven}},\ and\ \bibinfo {author} {\bibfnamefont {P.}~\bibnamefont {Kohli}},\ }\bibfield  {journal} {\bibinfo  {journal} {arXiv}\ }\href {https://doi.org/10.48550/arXiv.2310.05900} {10.48550/arXiv.2310.05900} (\bibinfo {year} {2023})\BibitemShut {NoStop}%
\bibitem [{\citenamefont {Higgott}(2022)}]{higgott_pymatching_2022}%
  \BibitemOpen
  \bibfield  {author} {\bibinfo {author} {\bibfnamefont {O.}~\bibnamefont {Higgott}},\ }\href {https://doi.org/10.1145/3505637} {\bibfield  {journal} {\bibinfo  {journal} {ACM Transactions on Quantum Computing}\ }\textbf {\bibinfo {volume} {3}},\ \bibinfo {pages} {16:1} (\bibinfo {year} {2022})}\BibitemShut {NoStop}%
\bibitem [{\citenamefont {Dennis}\ \emph {et~al.}(2002)\citenamefont {Dennis}, \citenamefont {Kitaev}, \citenamefont {Landahl},\ and\ \citenamefont {Preskill}}]{dennis_topological_2002}%
  \BibitemOpen
  \bibfield  {author} {\bibinfo {author} {\bibfnamefont {E.}~\bibnamefont {Dennis}}, \bibinfo {author} {\bibfnamefont {A.}~\bibnamefont {Kitaev}}, \bibinfo {author} {\bibfnamefont {A.}~\bibnamefont {Landahl}},\ and\ \bibinfo {author} {\bibfnamefont {J.}~\bibnamefont {Preskill}},\ }\href {https://doi.org/10.1063/1.1499754} {\bibfield  {journal} {\bibinfo  {journal} {Journal of Mathematical Physics}\ }\textbf {\bibinfo {volume} {43}},\ \bibinfo {pages} {4452} (\bibinfo {year} {2002})}\BibitemShut {NoStop}%
\bibitem [{\citenamefont {Chamberland}\ \emph {et~al.}(2020)\citenamefont {Chamberland}, \citenamefont {Zhu}, \citenamefont {Yoder}, \citenamefont {Hertzberg},\ and\ \citenamefont {Cross}}]{chamberland_topological_2020}%
  \BibitemOpen
  \bibfield  {author} {\bibinfo {author} {\bibfnamefont {C.}~\bibnamefont {Chamberland}}, \bibinfo {author} {\bibfnamefont {G.}~\bibnamefont {Zhu}}, \bibinfo {author} {\bibfnamefont {T.~J.}\ \bibnamefont {Yoder}}, \bibinfo {author} {\bibfnamefont {J.~B.}\ \bibnamefont {Hertzberg}},\ and\ \bibinfo {author} {\bibfnamefont {A.~W.}\ \bibnamefont {Cross}},\ }\href {https://doi.org/10.1103/PhysRevX.10.011022} {\bibfield  {journal} {\bibinfo  {journal} {Physical Review X}\ }\textbf {\bibinfo {volume} {10}},\ \bibinfo {pages} {011022} (\bibinfo {year} {2020})}\BibitemShut {NoStop}%
\bibitem [{\citenamefont {Nation}\ \emph {et~al.}(2021)\citenamefont {Nation}, \citenamefont {Paik}, \citenamefont {Cross},\ and\ \citenamefont {Nazario}}]{nation_ibm_2021}%
  \BibitemOpen
  \bibfield  {author} {\bibinfo {author} {\bibfnamefont {P.}~\bibnamefont {Nation}}, \bibinfo {author} {\bibfnamefont {H.}~\bibnamefont {Paik}}, \bibinfo {author} {\bibfnamefont {A.}~\bibnamefont {Cross}},\ and\ \bibinfo {author} {\bibfnamefont {Z.}~\bibnamefont {Nazario}},\ }\href {https://research.ibm.com/blog/heavy-hex-lattice} {\bibinfo {title} {The {IBM} {Quantum} heavy hex lattice}} (\bibinfo {year} {2021})\BibitemShut {NoStop}%
\bibitem [{\citenamefont {Sundaresan}\ \emph {et~al.}(2023)\citenamefont {Sundaresan}, \citenamefont {Yoder}, \citenamefont {Kim}, \citenamefont {Li}, \citenamefont {Chen}, \citenamefont {Harper}, \citenamefont {Thorbeck}, \citenamefont {Cross}, \citenamefont {Córcoles},\ and\ \citenamefont {Takita}}]{sundaresan_demonstrating_2023}%
  \BibitemOpen
  \bibfield  {author} {\bibinfo {author} {\bibfnamefont {N.}~\bibnamefont {Sundaresan}}, \bibinfo {author} {\bibfnamefont {T.~J.}\ \bibnamefont {Yoder}}, \bibinfo {author} {\bibfnamefont {Y.}~\bibnamefont {Kim}}, \bibinfo {author} {\bibfnamefont {M.}~\bibnamefont {Li}}, \bibinfo {author} {\bibfnamefont {E.~H.}\ \bibnamefont {Chen}}, \bibinfo {author} {\bibfnamefont {G.}~\bibnamefont {Harper}}, \bibinfo {author} {\bibfnamefont {T.}~\bibnamefont {Thorbeck}}, \bibinfo {author} {\bibfnamefont {A.~W.}\ \bibnamefont {Cross}}, \bibinfo {author} {\bibfnamefont {A.~D.}\ \bibnamefont {Córcoles}},\ and\ \bibinfo {author} {\bibfnamefont {M.}~\bibnamefont {Takita}},\ }\href {https://doi.org/10.1038/s41467-023-38247-5} {\bibfield  {journal} {\bibinfo  {journal} {Nature Communications}\ }\textbf {\bibinfo {volume} {14}},\ \bibinfo {pages} {2852} (\bibinfo {year} {2023})}\BibitemShut {NoStop}%
\bibitem [{\citenamefont {Skoric}\ \emph {et~al.}(2023)\citenamefont {Skoric}, \citenamefont {Browne}, \citenamefont {Barnes}, \citenamefont {Gillespie},\ and\ \citenamefont {Campbell}}]{skoric_parallel_2023}%
  \BibitemOpen
  \bibfield  {author} {\bibinfo {author} {\bibfnamefont {L.}~\bibnamefont {Skoric}}, \bibinfo {author} {\bibfnamefont {D.~E.}\ \bibnamefont {Browne}}, \bibinfo {author} {\bibfnamefont {K.~M.}\ \bibnamefont {Barnes}}, \bibinfo {author} {\bibfnamefont {N.~I.}\ \bibnamefont {Gillespie}},\ and\ \bibinfo {author} {\bibfnamefont {E.~T.}\ \bibnamefont {Campbell}},\ }\href {https://doi.org/10.1038/s41467-023-42482-1} {\bibfield  {journal} {\bibinfo  {journal} {Nature Communications}\ }\textbf {\bibinfo {volume} {14}},\ \bibinfo {pages} {7040} (\bibinfo {year} {2023})}\BibitemShut {NoStop}%
\bibitem [{\citenamefont {Battistel}\ \emph {et~al.}(2023)\citenamefont {Battistel}, \citenamefont {Chamberland}, \citenamefont {Johar}, \citenamefont {Overwater}, \citenamefont {Sebastiano}, \citenamefont {Skoric}, \citenamefont {Ueno},\ and\ \citenamefont {Usman}}]{battistel_real-time_2023}%
  \BibitemOpen
  \bibfield  {author} {\bibinfo {author} {\bibfnamefont {F.}~\bibnamefont {Battistel}}, \bibinfo {author} {\bibfnamefont {C.}~\bibnamefont {Chamberland}}, \bibinfo {author} {\bibfnamefont {K.}~\bibnamefont {Johar}}, \bibinfo {author} {\bibfnamefont {R.~W.~J.}\ \bibnamefont {Overwater}}, \bibinfo {author} {\bibfnamefont {F.}~\bibnamefont {Sebastiano}}, \bibinfo {author} {\bibfnamefont {L.}~\bibnamefont {Skoric}}, \bibinfo {author} {\bibfnamefont {Y.}~\bibnamefont {Ueno}},\ and\ \bibinfo {author} {\bibfnamefont {M.}~\bibnamefont {Usman}},\ }\href {https://doi.org/10.1088/2399-1984/aceba6} {\bibfield  {journal} {\bibinfo  {journal} {Nano Futures}\ }\textbf {\bibinfo {volume} {7}},\ \bibinfo {pages} {032003} (\bibinfo {year} {2023})}\BibitemShut {NoStop}%
\bibitem [{\citenamefont {Fowler}(2015)}]{fowler_minimum_2015}%
  \BibitemOpen
  \bibfield  {author} {\bibinfo {author} {\bibfnamefont {A.~G.}\ \bibnamefont {Fowler}},\ }\href {https://doi.org/10.48550/arXiv.1307.1740} {\bibfield  {journal} {\bibinfo  {journal} {Quantum Information \& Computation}\ }\textbf {\bibinfo {volume} {15}},\ \bibinfo {pages} {145} (\bibinfo {year} {2015})}\BibitemShut {NoStop}%
\bibitem [{\citenamefont {Acharya}\ \emph {et~al.}(2023)\citenamefont {Acharya}, \citenamefont {Aleiner}, \citenamefont {Allen}, \citenamefont {Andersen}, \citenamefont {Ansmann}, \citenamefont {Arute}, \citenamefont {Arya}, \citenamefont {Asfaw}, \citenamefont {Atalaya}, \citenamefont {Babbush}, \citenamefont {Bacon}, \citenamefont {Bardin}, \citenamefont {Basso}, \citenamefont {Bengtsson}, \citenamefont {Boixo}, \citenamefont {Bortoli}, \citenamefont {Bourassa}, \citenamefont {Bovaird}, \citenamefont {Brill}, \citenamefont {Broughton}, \citenamefont {Buckley}, \citenamefont {Buell}, \citenamefont {Burger}, \citenamefont {Burkett}, \citenamefont {Bushnell}, \citenamefont {Chen}, \citenamefont {Chen}, \citenamefont {Chiaro}, \citenamefont {Cogan}, \citenamefont {Collins}, \citenamefont {Conner}, \citenamefont {Courtney}, \citenamefont {Crook}, \citenamefont {Curtin}, \citenamefont {Debroy}, \citenamefont {Del Toro~Barba}, \citenamefont {Demura}, \citenamefont {Dunsworth}, \citenamefont {Eppens}, \citenamefont
  {Erickson}, \citenamefont {Faoro}, \citenamefont {Farhi}, \citenamefont {Fatemi}, \citenamefont {Flores~Burgos}, \citenamefont {Forati}, \citenamefont {Fowler}, \citenamefont {Foxen}, \citenamefont {Giang}, \citenamefont {Gidney}, \citenamefont {Gilboa}, \citenamefont {Giustina}, \citenamefont {Grajales~Dau}, \citenamefont {Gross}, \citenamefont {Habegger}, \citenamefont {Hamilton}, \citenamefont {Harrigan}, \citenamefont {Harrington}, \citenamefont {Higgott}, \citenamefont {Hilton}, \citenamefont {Hoffmann}, \citenamefont {Hong}, \citenamefont {Huang}, \citenamefont {Huff}, \citenamefont {Huggins}, \citenamefont {Ioffe}, \citenamefont {Isakov}, \citenamefont {Iveland}, \citenamefont {Jeffrey}, \citenamefont {Jiang}, \citenamefont {Jones}, \citenamefont {Juhas}, \citenamefont {Kafri}, \citenamefont {Kechedzhi}, \citenamefont {Kelly}, \citenamefont {Khattar}, \citenamefont {Khezri}, \citenamefont {Kieferová}, \citenamefont {Kim}, \citenamefont {Kitaev}, \citenamefont {Klimov}, \citenamefont {Klots},
  \citenamefont {Korotkov}, \citenamefont {Kostritsa}, \citenamefont {Kreikebaum}, \citenamefont {Landhuis}, \citenamefont {Laptev}, \citenamefont {Lau}, \citenamefont {Laws}, \citenamefont {Lee}, \citenamefont {Lee}, \citenamefont {Lester}, \citenamefont {Lill}, \citenamefont {Liu}, \citenamefont {Locharla}, \citenamefont {Lucero}, \citenamefont {Malone}, \citenamefont {Marshall}, \citenamefont {Martin}, \citenamefont {McClean}, \citenamefont {McCourt}, \citenamefont {McEwen}, \citenamefont {Megrant}, \citenamefont {Meurer~Costa}, \citenamefont {Mi}, \citenamefont {Miao}, \citenamefont {Mohseni}, \citenamefont {Montazeri}, \citenamefont {Morvan}, \citenamefont {Mount}, \citenamefont {Mruczkiewicz}, \citenamefont {Naaman}, \citenamefont {Neeley}, \citenamefont {Neill}, \citenamefont {Nersisyan}, \citenamefont {Neven}, \citenamefont {Newman}, \citenamefont {Ng}, \citenamefont {Nguyen}, \citenamefont {Nguyen}, \citenamefont {Niu}, \citenamefont {O’Brien}, \citenamefont {Opremcak}, \citenamefont {Platt},
  \citenamefont {Petukhov}, \citenamefont {Potter}, \citenamefont {Pryadko}, \citenamefont {Quintana}, \citenamefont {Roushan}, \citenamefont {Rubin}, \citenamefont {Saei}, \citenamefont {Sank}, \citenamefont {Sankaragomathi}, \citenamefont {Satzinger}, \citenamefont {Schurkus}, \citenamefont {Schuster}, \citenamefont {Shearn}, \citenamefont {Shorter}, \citenamefont {Shvarts}, \citenamefont {Skruzny}, \citenamefont {Smelyanskiy}, \citenamefont {Smith}, \citenamefont {Sterling}, \citenamefont {Strain}, \citenamefont {Szalay}, \citenamefont {Torres}, \citenamefont {Vidal}, \citenamefont {Villalonga}, \citenamefont {Vollgraff~Heidweiller}, \citenamefont {White}, \citenamefont {Xing}, \citenamefont {Yao}, \citenamefont {Yeh}, \citenamefont {Yoo}, \citenamefont {Young}, \citenamefont {Zalcman}, \citenamefont {Zhang}, \citenamefont {Zhu},\ and\ \citenamefont {{Google Quantum AI}}}]{acharya_suppressing_2023}%
  \BibitemOpen
  \bibfield  {author} {\bibinfo {author} {\bibfnamefont {R.}~\bibnamefont {Acharya}}, \bibinfo {author} {\bibfnamefont {I.}~\bibnamefont {Aleiner}}, \bibinfo {author} {\bibfnamefont {R.}~\bibnamefont {Allen}}, \bibinfo {author} {\bibfnamefont {T.~I.}\ \bibnamefont {Andersen}}, \bibinfo {author} {\bibfnamefont {M.}~\bibnamefont {Ansmann}}, \bibinfo {author} {\bibfnamefont {F.}~\bibnamefont {Arute}}, \bibinfo {author} {\bibfnamefont {K.}~\bibnamefont {Arya}}, \bibinfo {author} {\bibfnamefont {A.}~\bibnamefont {Asfaw}}, \bibinfo {author} {\bibfnamefont {J.}~\bibnamefont {Atalaya}}, \bibinfo {author} {\bibfnamefont {R.}~\bibnamefont {Babbush}}, \bibinfo {author} {\bibfnamefont {D.}~\bibnamefont {Bacon}}, \bibinfo {author} {\bibfnamefont {J.~C.}\ \bibnamefont {Bardin}}, \bibinfo {author} {\bibfnamefont {J.}~\bibnamefont {Basso}}, \bibinfo {author} {\bibfnamefont {A.}~\bibnamefont {Bengtsson}}, \bibinfo {author} {\bibfnamefont {S.}~\bibnamefont {Boixo}}, \bibinfo {author} {\bibfnamefont {G.}~\bibnamefont
  {Bortoli}}, \bibinfo {author} {\bibfnamefont {A.}~\bibnamefont {Bourassa}}, \bibinfo {author} {\bibfnamefont {J.}~\bibnamefont {Bovaird}}, \bibinfo {author} {\bibfnamefont {L.}~\bibnamefont {Brill}}, \bibinfo {author} {\bibfnamefont {M.}~\bibnamefont {Broughton}}, \bibinfo {author} {\bibfnamefont {B.~B.}\ \bibnamefont {Buckley}}, \bibinfo {author} {\bibfnamefont {D.~A.}\ \bibnamefont {Buell}}, \bibinfo {author} {\bibfnamefont {T.}~\bibnamefont {Burger}}, \bibinfo {author} {\bibfnamefont {B.}~\bibnamefont {Burkett}}, \bibinfo {author} {\bibfnamefont {N.}~\bibnamefont {Bushnell}}, \bibinfo {author} {\bibfnamefont {Y.}~\bibnamefont {Chen}}, \bibinfo {author} {\bibfnamefont {Z.}~\bibnamefont {Chen}}, \bibinfo {author} {\bibfnamefont {B.}~\bibnamefont {Chiaro}}, \bibinfo {author} {\bibfnamefont {J.}~\bibnamefont {Cogan}}, \bibinfo {author} {\bibfnamefont {R.}~\bibnamefont {Collins}}, \bibinfo {author} {\bibfnamefont {P.}~\bibnamefont {Conner}}, \bibinfo {author} {\bibfnamefont {W.}~\bibnamefont {Courtney}},
  \bibinfo {author} {\bibfnamefont {A.~L.}\ \bibnamefont {Crook}}, \bibinfo {author} {\bibfnamefont {B.}~\bibnamefont {Curtin}}, \bibinfo {author} {\bibfnamefont {D.~M.}\ \bibnamefont {Debroy}}, \bibinfo {author} {\bibfnamefont {A.}~\bibnamefont {Del Toro~Barba}}, \bibinfo {author} {\bibfnamefont {S.}~\bibnamefont {Demura}}, \bibinfo {author} {\bibfnamefont {A.}~\bibnamefont {Dunsworth}}, \bibinfo {author} {\bibfnamefont {D.}~\bibnamefont {Eppens}}, \bibinfo {author} {\bibfnamefont {C.}~\bibnamefont {Erickson}}, \bibinfo {author} {\bibfnamefont {L.}~\bibnamefont {Faoro}}, \bibinfo {author} {\bibfnamefont {E.}~\bibnamefont {Farhi}}, \bibinfo {author} {\bibfnamefont {R.}~\bibnamefont {Fatemi}}, \bibinfo {author} {\bibfnamefont {L.}~\bibnamefont {Flores~Burgos}}, \bibinfo {author} {\bibfnamefont {E.}~\bibnamefont {Forati}}, \bibinfo {author} {\bibfnamefont {A.~G.}\ \bibnamefont {Fowler}}, \bibinfo {author} {\bibfnamefont {B.}~\bibnamefont {Foxen}}, \bibinfo {author} {\bibfnamefont {W.}~\bibnamefont {Giang}},
  \bibinfo {author} {\bibfnamefont {C.}~\bibnamefont {Gidney}}, \bibinfo {author} {\bibfnamefont {D.}~\bibnamefont {Gilboa}}, \bibinfo {author} {\bibfnamefont {M.}~\bibnamefont {Giustina}}, \bibinfo {author} {\bibfnamefont {A.}~\bibnamefont {Grajales~Dau}}, \bibinfo {author} {\bibfnamefont {J.~A.}\ \bibnamefont {Gross}}, \bibinfo {author} {\bibfnamefont {S.}~\bibnamefont {Habegger}}, \bibinfo {author} {\bibfnamefont {M.~C.}\ \bibnamefont {Hamilton}}, \bibinfo {author} {\bibfnamefont {M.~P.}\ \bibnamefont {Harrigan}}, \bibinfo {author} {\bibfnamefont {S.~D.}\ \bibnamefont {Harrington}}, \bibinfo {author} {\bibfnamefont {O.}~\bibnamefont {Higgott}}, \bibinfo {author} {\bibfnamefont {J.}~\bibnamefont {Hilton}}, \bibinfo {author} {\bibfnamefont {M.}~\bibnamefont {Hoffmann}}, \bibinfo {author} {\bibfnamefont {S.}~\bibnamefont {Hong}}, \bibinfo {author} {\bibfnamefont {T.}~\bibnamefont {Huang}}, \bibinfo {author} {\bibfnamefont {A.}~\bibnamefont {Huff}}, \bibinfo {author} {\bibfnamefont {W.~J.}\ \bibnamefont
  {Huggins}}, \bibinfo {author} {\bibfnamefont {L.~B.}\ \bibnamefont {Ioffe}}, \bibinfo {author} {\bibfnamefont {S.~V.}\ \bibnamefont {Isakov}}, \bibinfo {author} {\bibfnamefont {J.}~\bibnamefont {Iveland}}, \bibinfo {author} {\bibfnamefont {E.}~\bibnamefont {Jeffrey}}, \bibinfo {author} {\bibfnamefont {Z.}~\bibnamefont {Jiang}}, \bibinfo {author} {\bibfnamefont {C.}~\bibnamefont {Jones}}, \bibinfo {author} {\bibfnamefont {P.}~\bibnamefont {Juhas}}, \bibinfo {author} {\bibfnamefont {D.}~\bibnamefont {Kafri}}, \bibinfo {author} {\bibfnamefont {K.}~\bibnamefont {Kechedzhi}}, \bibinfo {author} {\bibfnamefont {J.}~\bibnamefont {Kelly}}, \bibinfo {author} {\bibfnamefont {T.}~\bibnamefont {Khattar}}, \bibinfo {author} {\bibfnamefont {M.}~\bibnamefont {Khezri}}, \bibinfo {author} {\bibfnamefont {M.}~\bibnamefont {Kieferová}}, \bibinfo {author} {\bibfnamefont {S.}~\bibnamefont {Kim}}, \bibinfo {author} {\bibfnamefont {A.}~\bibnamefont {Kitaev}}, \bibinfo {author} {\bibfnamefont {P.~V.}\ \bibnamefont {Klimov}},
  \bibinfo {author} {\bibfnamefont {A.~R.}\ \bibnamefont {Klots}}, \bibinfo {author} {\bibfnamefont {A.~N.}\ \bibnamefont {Korotkov}}, \bibinfo {author} {\bibfnamefont {F.}~\bibnamefont {Kostritsa}}, \bibinfo {author} {\bibfnamefont {J.~M.}\ \bibnamefont {Kreikebaum}}, \bibinfo {author} {\bibfnamefont {D.}~\bibnamefont {Landhuis}}, \bibinfo {author} {\bibfnamefont {P.}~\bibnamefont {Laptev}}, \bibinfo {author} {\bibfnamefont {K.-M.}\ \bibnamefont {Lau}}, \bibinfo {author} {\bibfnamefont {L.}~\bibnamefont {Laws}}, \bibinfo {author} {\bibfnamefont {J.}~\bibnamefont {Lee}}, \bibinfo {author} {\bibfnamefont {K.}~\bibnamefont {Lee}}, \bibinfo {author} {\bibfnamefont {B.~J.}\ \bibnamefont {Lester}}, \bibinfo {author} {\bibfnamefont {A.}~\bibnamefont {Lill}}, \bibinfo {author} {\bibfnamefont {W.}~\bibnamefont {Liu}}, \bibinfo {author} {\bibfnamefont {A.}~\bibnamefont {Locharla}}, \bibinfo {author} {\bibfnamefont {E.}~\bibnamefont {Lucero}}, \bibinfo {author} {\bibfnamefont {F.~D.}\ \bibnamefont {Malone}}, \bibinfo
  {author} {\bibfnamefont {J.}~\bibnamefont {Marshall}}, \bibinfo {author} {\bibfnamefont {O.}~\bibnamefont {Martin}}, \bibinfo {author} {\bibfnamefont {J.~R.}\ \bibnamefont {McClean}}, \bibinfo {author} {\bibfnamefont {T.}~\bibnamefont {McCourt}}, \bibinfo {author} {\bibfnamefont {M.}~\bibnamefont {McEwen}}, \bibinfo {author} {\bibfnamefont {A.}~\bibnamefont {Megrant}}, \bibinfo {author} {\bibfnamefont {B.}~\bibnamefont {Meurer~Costa}}, \bibinfo {author} {\bibfnamefont {X.}~\bibnamefont {Mi}}, \bibinfo {author} {\bibfnamefont {K.~C.}\ \bibnamefont {Miao}}, \bibinfo {author} {\bibfnamefont {M.}~\bibnamefont {Mohseni}}, \bibinfo {author} {\bibfnamefont {S.}~\bibnamefont {Montazeri}}, \bibinfo {author} {\bibfnamefont {A.}~\bibnamefont {Morvan}}, \bibinfo {author} {\bibfnamefont {E.}~\bibnamefont {Mount}}, \bibinfo {author} {\bibfnamefont {W.}~\bibnamefont {Mruczkiewicz}}, \bibinfo {author} {\bibfnamefont {O.}~\bibnamefont {Naaman}}, \bibinfo {author} {\bibfnamefont {M.}~\bibnamefont {Neeley}}, \bibinfo {author}
  {\bibfnamefont {C.}~\bibnamefont {Neill}}, \bibinfo {author} {\bibfnamefont {A.}~\bibnamefont {Nersisyan}}, \bibinfo {author} {\bibfnamefont {H.}~\bibnamefont {Neven}}, \bibinfo {author} {\bibfnamefont {M.}~\bibnamefont {Newman}}, \bibinfo {author} {\bibfnamefont {J.~H.}\ \bibnamefont {Ng}}, \bibinfo {author} {\bibfnamefont {A.}~\bibnamefont {Nguyen}}, \bibinfo {author} {\bibfnamefont {M.}~\bibnamefont {Nguyen}}, \bibinfo {author} {\bibfnamefont {M.~Y.}\ \bibnamefont {Niu}}, \bibinfo {author} {\bibfnamefont {T.~E.}\ \bibnamefont {O’Brien}}, \bibinfo {author} {\bibfnamefont {A.}~\bibnamefont {Opremcak}}, \bibinfo {author} {\bibfnamefont {J.}~\bibnamefont {Platt}}, \bibinfo {author} {\bibfnamefont {A.}~\bibnamefont {Petukhov}}, \bibinfo {author} {\bibfnamefont {R.}~\bibnamefont {Potter}}, \bibinfo {author} {\bibfnamefont {L.~P.}\ \bibnamefont {Pryadko}}, \bibinfo {author} {\bibfnamefont {C.}~\bibnamefont {Quintana}}, \bibinfo {author} {\bibfnamefont {P.}~\bibnamefont {Roushan}}, \bibinfo {author}
  {\bibfnamefont {N.~C.}\ \bibnamefont {Rubin}}, \bibinfo {author} {\bibfnamefont {N.}~\bibnamefont {Saei}}, \bibinfo {author} {\bibfnamefont {D.}~\bibnamefont {Sank}}, \bibinfo {author} {\bibfnamefont {K.}~\bibnamefont {Sankaragomathi}}, \bibinfo {author} {\bibfnamefont {K.~J.}\ \bibnamefont {Satzinger}}, \bibinfo {author} {\bibfnamefont {H.~F.}\ \bibnamefont {Schurkus}}, \bibinfo {author} {\bibfnamefont {C.}~\bibnamefont {Schuster}}, \bibinfo {author} {\bibfnamefont {M.~J.}\ \bibnamefont {Shearn}}, \bibinfo {author} {\bibfnamefont {A.}~\bibnamefont {Shorter}}, \bibinfo {author} {\bibfnamefont {V.}~\bibnamefont {Shvarts}}, \bibinfo {author} {\bibfnamefont {J.}~\bibnamefont {Skruzny}}, \bibinfo {author} {\bibfnamefont {V.}~\bibnamefont {Smelyanskiy}}, \bibinfo {author} {\bibfnamefont {W.~C.}\ \bibnamefont {Smith}}, \bibinfo {author} {\bibfnamefont {G.}~\bibnamefont {Sterling}}, \bibinfo {author} {\bibfnamefont {D.}~\bibnamefont {Strain}}, \bibinfo {author} {\bibfnamefont {M.}~\bibnamefont {Szalay}}, \bibinfo
  {author} {\bibfnamefont {A.}~\bibnamefont {Torres}}, \bibinfo {author} {\bibfnamefont {G.}~\bibnamefont {Vidal}}, \bibinfo {author} {\bibfnamefont {B.}~\bibnamefont {Villalonga}}, \bibinfo {author} {\bibfnamefont {C.}~\bibnamefont {Vollgraff~Heidweiller}}, \bibinfo {author} {\bibfnamefont {T.}~\bibnamefont {White}}, \bibinfo {author} {\bibfnamefont {C.}~\bibnamefont {Xing}}, \bibinfo {author} {\bibfnamefont {Z.~J.}\ \bibnamefont {Yao}}, \bibinfo {author} {\bibfnamefont {P.}~\bibnamefont {Yeh}}, \bibinfo {author} {\bibfnamefont {J.}~\bibnamefont {Yoo}}, \bibinfo {author} {\bibfnamefont {G.}~\bibnamefont {Young}}, \bibinfo {author} {\bibfnamefont {A.}~\bibnamefont {Zalcman}}, \bibinfo {author} {\bibfnamefont {Y.}~\bibnamefont {Zhang}}, \bibinfo {author} {\bibfnamefont {N.}~\bibnamefont {Zhu}},\ and\ \bibinfo {author} {\bibnamefont {{Google Quantum AI}}},\ }\href {https://doi.org/10.1038/s41586-022-05434-1} {\bibfield  {journal} {\bibinfo  {journal} {Nature}\ }\textbf {\bibinfo {volume} {614}},\ \bibinfo
  {pages} {676} (\bibinfo {year} {2023})}\BibitemShut {NoStop}%
\bibitem [{\citenamefont {Fowler}\ \emph {et~al.}(2012{\natexlab{b}})\citenamefont {Fowler}, \citenamefont {Whiteside},\ and\ \citenamefont {Hollenberg}}]{fowler_towards_2012}%
  \BibitemOpen
  \bibfield  {author} {\bibinfo {author} {\bibfnamefont {A.~G.}\ \bibnamefont {Fowler}}, \bibinfo {author} {\bibfnamefont {A.~C.}\ \bibnamefont {Whiteside}},\ and\ \bibinfo {author} {\bibfnamefont {L.~C.~L.}\ \bibnamefont {Hollenberg}},\ }\href {https://doi.org/10.1103/PhysRevLett.108.180501} {\bibfield  {journal} {\bibinfo  {journal} {Physical Review Letters}\ }\textbf {\bibinfo {volume} {108}},\ \bibinfo {pages} {180501} (\bibinfo {year} {2012}{\natexlab{b}})}\BibitemShut {NoStop}%
\bibitem [{\citenamefont {Wang}\ \emph {et~al.}(2010)\citenamefont {Wang}, \citenamefont {Fowler}, \citenamefont {Stephens},\ and\ \citenamefont {Hollenberg}}]{wang_threshold_2010}%
  \BibitemOpen
  \bibfield  {author} {\bibinfo {author} {\bibfnamefont {D.~S.}\ \bibnamefont {Wang}}, \bibinfo {author} {\bibfnamefont {A.~G.}\ \bibnamefont {Fowler}}, \bibinfo {author} {\bibfnamefont {A.~M.}\ \bibnamefont {Stephens}},\ and\ \bibinfo {author} {\bibfnamefont {L.~C.~L.}\ \bibnamefont {Hollenberg}},\ }\href {https://doi.org/10.48550/arXiv.0905.0531} {\bibfield  {journal} {\bibinfo  {journal} {Quantum Information \& Computation}\ }\textbf {\bibinfo {volume} {10}},\ \bibinfo {pages} {456} (\bibinfo {year} {2010})}\BibitemShut {NoStop}%
\bibitem [{\citenamefont {Delfosse}\ \emph {et~al.}(2023)\citenamefont {Delfosse}, \citenamefont {Paz}, \citenamefont {Vaschillo},\ and\ \citenamefont {Svore}}]{delfosse_how_2023}%
  \BibitemOpen
  \bibfield  {author} {\bibinfo {author} {\bibfnamefont {N.}~\bibnamefont {Delfosse}}, \bibinfo {author} {\bibfnamefont {A.}~\bibnamefont {Paz}}, \bibinfo {author} {\bibfnamefont {A.}~\bibnamefont {Vaschillo}},\ and\ \bibinfo {author} {\bibfnamefont {K.~M.}\ \bibnamefont {Svore}},\ }\bibfield  {journal} {\bibinfo  {journal} {arXiv}\ }\href {https://doi.org/10.48550/arXiv.2310.15313} {10.48550/arXiv.2310.15313} (\bibinfo {year} {2023})\BibitemShut {NoStop}%
\bibitem [{\citenamefont {Torlai}\ and\ \citenamefont {Melko}(2017)}]{torlai_neural_2017}%
  \BibitemOpen
  \bibfield  {author} {\bibinfo {author} {\bibfnamefont {G.}~\bibnamefont {Torlai}}\ and\ \bibinfo {author} {\bibfnamefont {R.~G.}\ \bibnamefont {Melko}},\ }\href {https://doi.org/10.1103/PhysRevLett.119.030501} {\bibfield  {journal} {\bibinfo  {journal} {Physical Review Letters}\ }\textbf {\bibinfo {volume} {119}},\ \bibinfo {pages} {030501} (\bibinfo {year} {2017})}\BibitemShut {NoStop}%
\bibitem [{\citenamefont {Chamberland}\ and\ \citenamefont {Ronagh}(2018)}]{chamberland_deep_2018}%
  \BibitemOpen
  \bibfield  {author} {\bibinfo {author} {\bibfnamefont {C.}~\bibnamefont {Chamberland}}\ and\ \bibinfo {author} {\bibfnamefont {P.}~\bibnamefont {Ronagh}},\ }\href {https://doi.org/10.1088/2058-9565/aad1f7} {\bibfield  {journal} {\bibinfo  {journal} {Quantum Science and Technology}\ }\textbf {\bibinfo {volume} {3}},\ \bibinfo {pages} {044002} (\bibinfo {year} {2018})}\BibitemShut {NoStop}%
\bibitem [{\citenamefont {Krastanov}\ and\ \citenamefont {Jiang}(2017)}]{krastanov_deep_2017}%
  \BibitemOpen
  \bibfield  {author} {\bibinfo {author} {\bibfnamefont {S.}~\bibnamefont {Krastanov}}\ and\ \bibinfo {author} {\bibfnamefont {L.}~\bibnamefont {Jiang}},\ }\href {https://doi.org/10.1038/s41598-017-11266-1} {\bibfield  {journal} {\bibinfo  {journal} {Scientific Reports}\ }\textbf {\bibinfo {volume} {7}},\ \bibinfo {pages} {11003} (\bibinfo {year} {2017})}\BibitemShut {NoStop}%
\bibitem [{\citenamefont {Baireuther}\ \emph {et~al.}(2018)\citenamefont {Baireuther}, \citenamefont {O'Brien}, \citenamefont {Tarasinski},\ and\ \citenamefont {Beenakker}}]{baireuther_machine-learning-assisted_2018}%
  \BibitemOpen
  \bibfield  {author} {\bibinfo {author} {\bibfnamefont {P.}~\bibnamefont {Baireuther}}, \bibinfo {author} {\bibfnamefont {T.~E.}\ \bibnamefont {O'Brien}}, \bibinfo {author} {\bibfnamefont {B.}~\bibnamefont {Tarasinski}},\ and\ \bibinfo {author} {\bibfnamefont {C.~W.~J.}\ \bibnamefont {Beenakker}},\ }\href {https://doi.org/10.22331/q-2018-01-29-48} {\bibfield  {journal} {\bibinfo  {journal} {Quantum}\ }\textbf {\bibinfo {volume} {2}},\ \bibinfo {pages} {48} (\bibinfo {year} {2018})}\BibitemShut {NoStop}%
\bibitem [{\citenamefont {Davaasuren}\ \emph {et~al.}(2020)\citenamefont {Davaasuren}, \citenamefont {Suzuki}, \citenamefont {Fujii},\ and\ \citenamefont {Koashi}}]{davaasuren_general_2020}%
  \BibitemOpen
  \bibfield  {author} {\bibinfo {author} {\bibfnamefont {A.}~\bibnamefont {Davaasuren}}, \bibinfo {author} {\bibfnamefont {Y.}~\bibnamefont {Suzuki}}, \bibinfo {author} {\bibfnamefont {K.}~\bibnamefont {Fujii}},\ and\ \bibinfo {author} {\bibfnamefont {M.}~\bibnamefont {Koashi}},\ }\href {https://doi.org/10.1103/PhysRevResearch.2.033399} {\bibfield  {journal} {\bibinfo  {journal} {Physical Review Research}\ }\textbf {\bibinfo {volume} {2}},\ \bibinfo {pages} {033399} (\bibinfo {year} {2020})}\BibitemShut {NoStop}%
\bibitem [{\citenamefont {Bhoumik}\ \emph {et~al.}(2022)\citenamefont {Bhoumik}, \citenamefont {Majumdar}, \citenamefont {Madan}, \citenamefont {Vinayagamurthy}, \citenamefont {Raghunathan},\ and\ \citenamefont {Sur-Kolay}}]{bhoumik_efficient_2022}%
  \BibitemOpen
  \bibfield  {author} {\bibinfo {author} {\bibfnamefont {D.}~\bibnamefont {Bhoumik}}, \bibinfo {author} {\bibfnamefont {R.}~\bibnamefont {Majumdar}}, \bibinfo {author} {\bibfnamefont {D.}~\bibnamefont {Madan}}, \bibinfo {author} {\bibfnamefont {D.}~\bibnamefont {Vinayagamurthy}}, \bibinfo {author} {\bibfnamefont {S.}~\bibnamefont {Raghunathan}},\ and\ \bibinfo {author} {\bibfnamefont {S.}~\bibnamefont {Sur-Kolay}},\ }\bibfield  {journal} {\bibinfo  {journal} {arXiv}\ }\href {https://doi.org/10.48550/arXiv.2210.09730} {10.48550/arXiv.2210.09730} (\bibinfo {year} {2022})\BibitemShut {NoStop}%
\bibitem [{\citenamefont {Li}\ \emph {et~al.}(2023)\citenamefont {Li}, \citenamefont {Li}, \citenamefont {Gan},\ and\ \citenamefont {Ma}}]{li_convolutional-neural-network-based_2023}%
  \BibitemOpen
  \bibfield  {author} {\bibinfo {author} {\bibfnamefont {A.}~\bibnamefont {Li}}, \bibinfo {author} {\bibfnamefont {F.}~\bibnamefont {Li}}, \bibinfo {author} {\bibfnamefont {Q.}~\bibnamefont {Gan}},\ and\ \bibinfo {author} {\bibfnamefont {H.}~\bibnamefont {Ma}},\ }\href {https://doi.org/10.3390/app13179689} {\bibfield  {journal} {\bibinfo  {journal} {Applied Sciences}\ }\textbf {\bibinfo {volume} {13}},\ \bibinfo {pages} {9689} (\bibinfo {year} {2023})}\BibitemShut {NoStop}%
\bibitem [{\citenamefont {Abadi}\ \emph {et~al.}(2016)\citenamefont {Abadi}, \citenamefont {Barham}, \citenamefont {Chen}, \citenamefont {Chen}, \citenamefont {Davis}, \citenamefont {Dean}, \citenamefont {Devin}, \citenamefont {Ghemawat}, \citenamefont {Irving}, \citenamefont {Isard}, \citenamefont {Kudlur}, \citenamefont {Levenberg}, \citenamefont {Monga}, \citenamefont {Moore}, \citenamefont {Murray}, \citenamefont {Steiner}, \citenamefont {Tucker}, \citenamefont {Vasudevan}, \citenamefont {Warden}, \citenamefont {Wicke}, \citenamefont {Yu},\ and\ \citenamefont {Zheng}}]{abadi_tensorflow_2016}%
  \BibitemOpen
  \bibfield  {author} {\bibinfo {author} {\bibfnamefont {M.}~\bibnamefont {Abadi}}, \bibinfo {author} {\bibfnamefont {P.}~\bibnamefont {Barham}}, \bibinfo {author} {\bibfnamefont {J.}~\bibnamefont {Chen}}, \bibinfo {author} {\bibfnamefont {Z.}~\bibnamefont {Chen}}, \bibinfo {author} {\bibfnamefont {A.}~\bibnamefont {Davis}}, \bibinfo {author} {\bibfnamefont {J.}~\bibnamefont {Dean}}, \bibinfo {author} {\bibfnamefont {M.}~\bibnamefont {Devin}}, \bibinfo {author} {\bibfnamefont {S.}~\bibnamefont {Ghemawat}}, \bibinfo {author} {\bibfnamefont {G.}~\bibnamefont {Irving}}, \bibinfo {author} {\bibfnamefont {M.}~\bibnamefont {Isard}}, \bibinfo {author} {\bibfnamefont {M.}~\bibnamefont {Kudlur}}, \bibinfo {author} {\bibfnamefont {J.}~\bibnamefont {Levenberg}}, \bibinfo {author} {\bibfnamefont {R.}~\bibnamefont {Monga}}, \bibinfo {author} {\bibfnamefont {S.}~\bibnamefont {Moore}}, \bibinfo {author} {\bibfnamefont {D.~G.}\ \bibnamefont {Murray}}, \bibinfo {author} {\bibfnamefont {B.}~\bibnamefont {Steiner}}, \bibinfo
  {author} {\bibfnamefont {P.}~\bibnamefont {Tucker}}, \bibinfo {author} {\bibfnamefont {V.}~\bibnamefont {Vasudevan}}, \bibinfo {author} {\bibfnamefont {P.}~\bibnamefont {Warden}}, \bibinfo {author} {\bibfnamefont {M.}~\bibnamefont {Wicke}}, \bibinfo {author} {\bibfnamefont {Y.}~\bibnamefont {Yu}},\ and\ \bibinfo {author} {\bibfnamefont {X.}~\bibnamefont {Zheng}},\ }in\ \href {https://doi.org/10.48550/arXiv.1605.08695} {\emph {\bibinfo {booktitle} {Proceedings of the 12th {USENIX} conference on {Operating} {Systems} {Design} and {Implementation}}}},\ \bibinfo {series and number} {{OSDI}'16}\ (\bibinfo  {publisher} {USENIX Association},\ \bibinfo {address} {USA},\ \bibinfo {year} {2016})\ pp.\ \bibinfo {pages} {265--283}\BibitemShut {NoStop}%
\bibitem [{\citenamefont {IBM}(2022)}]{ibm_ibm_2022}%
  \BibitemOpen
  \bibfield  {author} {\bibinfo {author} {\bibnamefont {IBM}},\ }\href {https://quantum-computing.ibm.com/} {\bibinfo {title} {{IBM} {Quantum}}} (\bibinfo {year} {2022})\BibitemShut {NoStop}%
\bibitem [{\citenamefont {Chamberland}\ \emph {et~al.}(2023)\citenamefont {Chamberland}, \citenamefont {Goncalves}, \citenamefont {Sivarajah}, \citenamefont {Peterson},\ and\ \citenamefont {Grimberg}}]{chamberland_techniques_2023}%
  \BibitemOpen
  \bibfield  {author} {\bibinfo {author} {\bibfnamefont {C.}~\bibnamefont {Chamberland}}, \bibinfo {author} {\bibfnamefont {L.}~\bibnamefont {Goncalves}}, \bibinfo {author} {\bibfnamefont {P.}~\bibnamefont {Sivarajah}}, \bibinfo {author} {\bibfnamefont {E.}~\bibnamefont {Peterson}},\ and\ \bibinfo {author} {\bibfnamefont {S.}~\bibnamefont {Grimberg}},\ }\href {https://doi.org/10.1088/2058-9565/ace64d} {\bibfield  {journal} {\bibinfo  {journal} {Quantum Science and Technology}\ }\textbf {\bibinfo {volume} {8}},\ \bibinfo {pages} {045011} (\bibinfo {year} {2023})}\BibitemShut {NoStop}%
\bibitem [{\citenamefont {Ueno}\ \emph {et~al.}(2022)\citenamefont {Ueno}, \citenamefont {Kondo}, \citenamefont {Tanaka}, \citenamefont {Suzuki},\ and\ \citenamefont {Tabuchi}}]{ueno_neo-qec_2022}%
  \BibitemOpen
  \bibfield  {author} {\bibinfo {author} {\bibfnamefont {Y.}~\bibnamefont {Ueno}}, \bibinfo {author} {\bibfnamefont {M.}~\bibnamefont {Kondo}}, \bibinfo {author} {\bibfnamefont {M.}~\bibnamefont {Tanaka}}, \bibinfo {author} {\bibfnamefont {Y.}~\bibnamefont {Suzuki}},\ and\ \bibinfo {author} {\bibfnamefont {Y.}~\bibnamefont {Tabuchi}},\ }\bibfield  {journal} {\bibinfo  {journal} {arXiv}\ }\href {https://doi.org/10.48550/arXiv.2208.05758} {10.48550/arXiv.2208.05758} (\bibinfo {year} {2022})\BibitemShut {NoStop}%
\bibitem [{\citenamefont {Tomita}\ and\ \citenamefont {Svore}(2014)}]{tomita_low-distance_2014}%
  \BibitemOpen
  \bibfield  {author} {\bibinfo {author} {\bibfnamefont {Y.}~\bibnamefont {Tomita}}\ and\ \bibinfo {author} {\bibfnamefont {K.~M.}\ \bibnamefont {Svore}},\ }\href {https://doi.org/10.1103/PhysRevA.90.062320} {\bibfield  {journal} {\bibinfo  {journal} {Physical Review A}\ }\textbf {\bibinfo {volume} {90}},\ \bibinfo {pages} {062320} (\bibinfo {year} {2014})}\BibitemShut {NoStop}%
\bibitem [{\citenamefont {Chen}\ \emph {et~al.}(2022)\citenamefont {Chen}, \citenamefont {Yoder}, \citenamefont {Kim}, \citenamefont {Sundaresan}, \citenamefont {Srinivasan}, \citenamefont {Li}, \citenamefont {Córcoles}, \citenamefont {Cross},\ and\ \citenamefont {Takita}}]{chen_calibrated_2022}%
  \BibitemOpen
  \bibfield  {author} {\bibinfo {author} {\bibfnamefont {E.~H.}\ \bibnamefont {Chen}}, \bibinfo {author} {\bibfnamefont {T.~J.}\ \bibnamefont {Yoder}}, \bibinfo {author} {\bibfnamefont {Y.}~\bibnamefont {Kim}}, \bibinfo {author} {\bibfnamefont {N.}~\bibnamefont {Sundaresan}}, \bibinfo {author} {\bibfnamefont {S.}~\bibnamefont {Srinivasan}}, \bibinfo {author} {\bibfnamefont {M.}~\bibnamefont {Li}}, \bibinfo {author} {\bibfnamefont {A.~D.}\ \bibnamefont {Córcoles}}, \bibinfo {author} {\bibfnamefont {A.~W.}\ \bibnamefont {Cross}},\ and\ \bibinfo {author} {\bibfnamefont {M.}~\bibnamefont {Takita}},\ }\href {https://doi.org/10.1103/PhysRevLett.128.110504} {\bibfield  {journal} {\bibinfo  {journal} {Physical Review Letters}\ }\textbf {\bibinfo {volume} {128}},\ \bibinfo {pages} {110504} (\bibinfo {year} {2022})}\BibitemShut {NoStop}%
\end{thebibliography}%


\clearpage
\newpage

\onecolumngrid
\begin{center}

\large{\textbf{Supplementary Material for}}\\\vspace{5pt}
\large{\textbf{``Artificial Neural Network Syndrome Decoding on IBM Quantum Processors"} }

\end{center}

\vspace{30pt}
\twocolumngrid
\normalsize

\renewcommand{\figurename}{SUPPLEMENTARY FIG.}
\renewcommand{\tablename}{SUPPLEMENTARY TABLE.}
\renewcommand{\thefigure}{S\arabic{figure}}
\renewcommand{\theequation}{S\arabic{equation}}
\renewcommand{\thetable}{S\arabic{equation}}
\renewcommand{\thesection}{S\arabic{section}}

\setcounter{figure}{0}

\section{Heavy Hexagon Adjustment}\label{sec:adjustment}
Across the structure of the HH code, qubits are labelled as either data, flag or measurement qubits. These different qubit types are what facilitate the locating of errors in the HH code. These form the basis of the stabiliser formalism for QEC codes.
Although IBM quantum processing devices have been developed for some years, the HH code which directly corresponds to the physical layout has not been discussed often, with only a few current works directly implementing the adjusted HH structure on superconducting transmon qubits \cite{sundaresan_demonstrating_2023, chen_calibrated_2022}.
Within the main text, it is stated that the HH boundary optimisation was not included when IBM physically realised their quantum devices.

\begin{figure}[h]
    \centering
    \includegraphics[width=\linewidth]{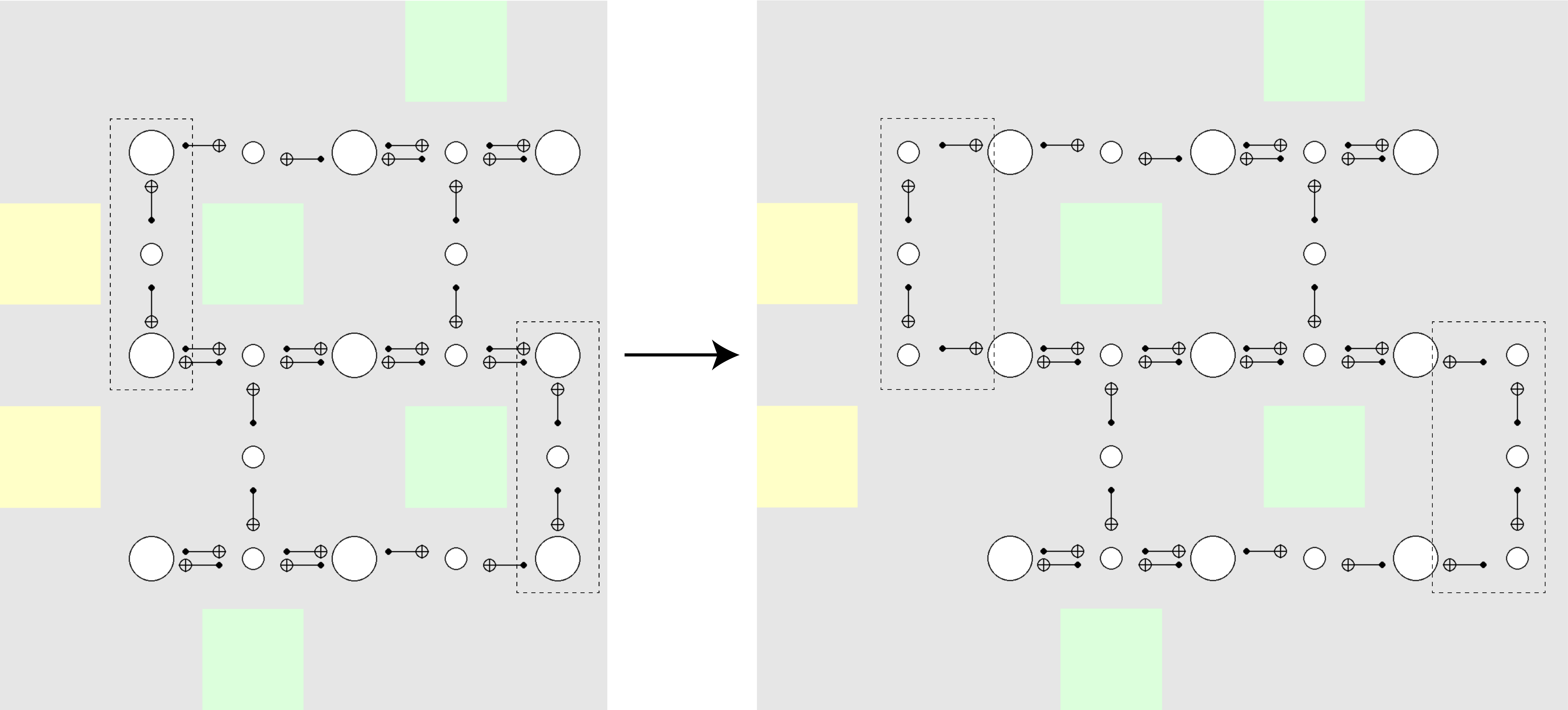}
    \caption{\textbf{Heavy Hexagon Boundary Adjustment.} The adjustment is made from the original HH QEC code structure (left), to the HH QEC code structure which fits to current IBM devices (right). The dashed box highlights the difference in boundary between the un-adjusted and adjusted codes. Circles represent qubits in the lattice, with larger and smaller being data and ancilla respectively. Yellow ($X$) and green ($Z$) squares refer to the stabilisers as measured by the flag and measurement qubits in the lattice. CNOTs are drawn on the lattice with corresponding directions for acts and controls.}
    \label{fig:adjustment}
\end{figure}

\begin{figure}[t]
    \centering
    \includegraphics[width=\linewidth]{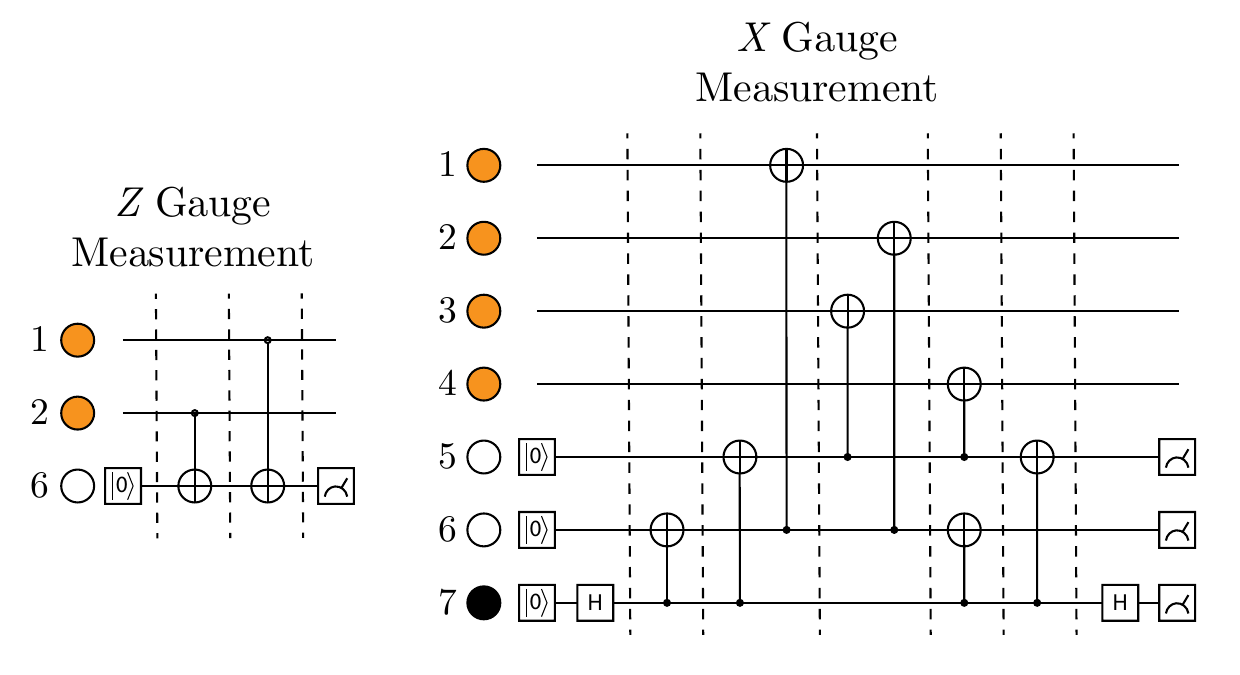}
    \caption{\textbf{Heavy Hexagon Gauge Measurement Circuits.} These two illustrated circuits describe the gauge operator measurements within the HH lattice. Orange circles represent data qubits, white circles represent flag qubits and black circles represent measurement qubits. These are numbered to match the numbering of qubits in Figure 1 of the main text. Flag/measurement qubits are initialised and measured at the beginning and final timesteps with CNOTs connecting the qubits at corresponding timesteps.}
    \label{fig:HHcircuits}
\end{figure}

In the Figure \ref{fig:adjustment}, the boundary optimisation shown on the left is removed on the right. The structure shown on the right hand side is physically implementable on IBM devices.
Within the adjusted HH lattice on the right of Figure \ref{fig:adjustment}, there are three types of stabiliser generator; the $X$-type Bacon-Shor style operators;
$$ S_X = \prod_n X_{n,j} X_{n,j+1} $$
the weight-four $Z$-type plaquette operators, found in the bulk;
$$ S_Z = Z_{i,j}Z_{i+1,j}Z_{i,j+1}Z_{i+1,j+1} $$
and the weight-two $Z$-type edge operators;
$$ S_Z = Z_{2m,1}Z_{2m+1,1}, \ Z_{2m-1,d}Z_{2m,d} $$
where $i,j \in \mathbb{N} \leq d-1$, $m \in \mathbb{N} \leq \frac{d-1}{2}$ and $n \in \mathbb{N} \leq d$, and $i+j = $ even in the second set. Here, $i,j$ refer to the lattice of data qubits, with $i$ as rows and $j$ as columns.
The stabiliser group, as used in QEC codes, is sufficiently defined by the stabiliser generators which form the entire group after all multiple combinations.
Given the boundary conditions of the device, the edge operators are found along the top and bottom of the lattice when arranged in the alignment of Figure \ref{fig:adjustment}. This is to ensure that operators do not act on non-existent qubits.
The result of measurement of the stabilisers across the lattice is the syndrome measurement. 
These generators mutually commute, allowing for their collective simultaneous measurement. 
Given that there are many ancillary qubits on the lattice, gauge operators are defined to localised areas to measure the local parity, and the stabilisers of each kind measure the parity of gauge operators of each kind. The gauge operators are defined as;
\begin{align*}
    G_X = \ &X_{i,j} X_{i+1,j} X_{i, j+1} X_{i+1, j+1}, \\ &X_{1, 2m-1} X_{1, 2m}, \ X_{d, 2m} X_{d, 2m+1}
\end{align*}
and;
$$ G_Z = Z_{i,j}Z_{i+1,j} $$
for $X$ and $Z$ gauge operators respectively, where $i,j \in \mathbb{N} \leq d$, $m \in \mathbb{N} \leq \frac{d-1}{2}$. A constraint of $i+j = $ odd must be used for the first term in the $X$ gauge operator set.
The measurements of these gauge operators and hence stabilisers can be facilitated by the gauge operator circuit diagrams illustrated in Figure \ref{fig:HHcircuits}.

Figure \ref{fig:HHd5error} illustrates the overall layout of a $d=5$ adjusted HH code with some data errors and corresponding stabiliser measurements.

\begin{figure*}[t]
    \centering
    \includegraphics[width=\linewidth]{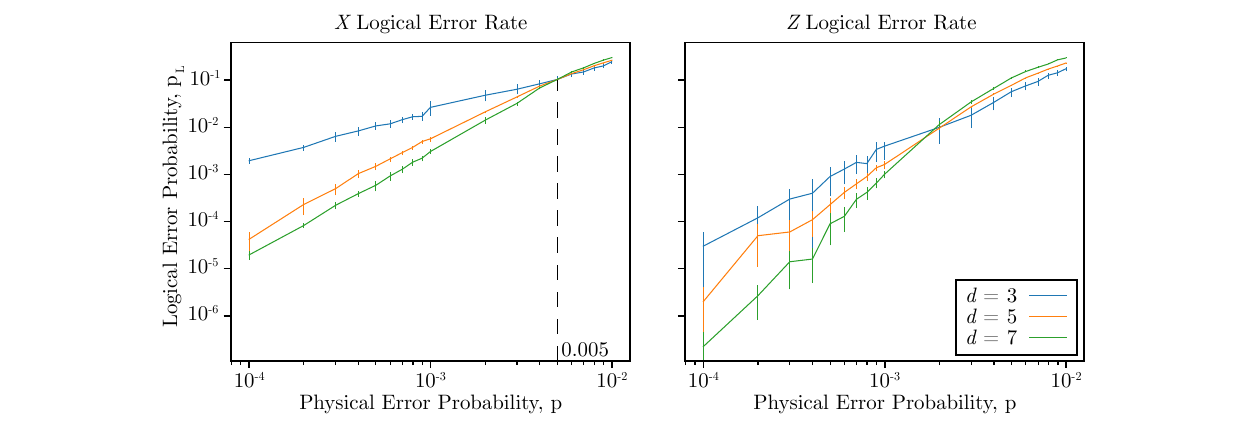}
    \caption{\textbf{Benchmarking of the original Heavy Hexagon Code with MWPM.} Both the threshold and psuedo-threshold for $X$ logical errors (left) and $Z$ logical errors (right) for the adjusted HH code are shown; decoded by MWPM as implemeted by PyMatching. Error bars are assigned with a probit corresponding to $97.5\%$. }
    \label{fig:HHMWPM}
\end{figure*}

\begin{figure}[h]
    \centering
    \includegraphics[width=0.9\linewidth]{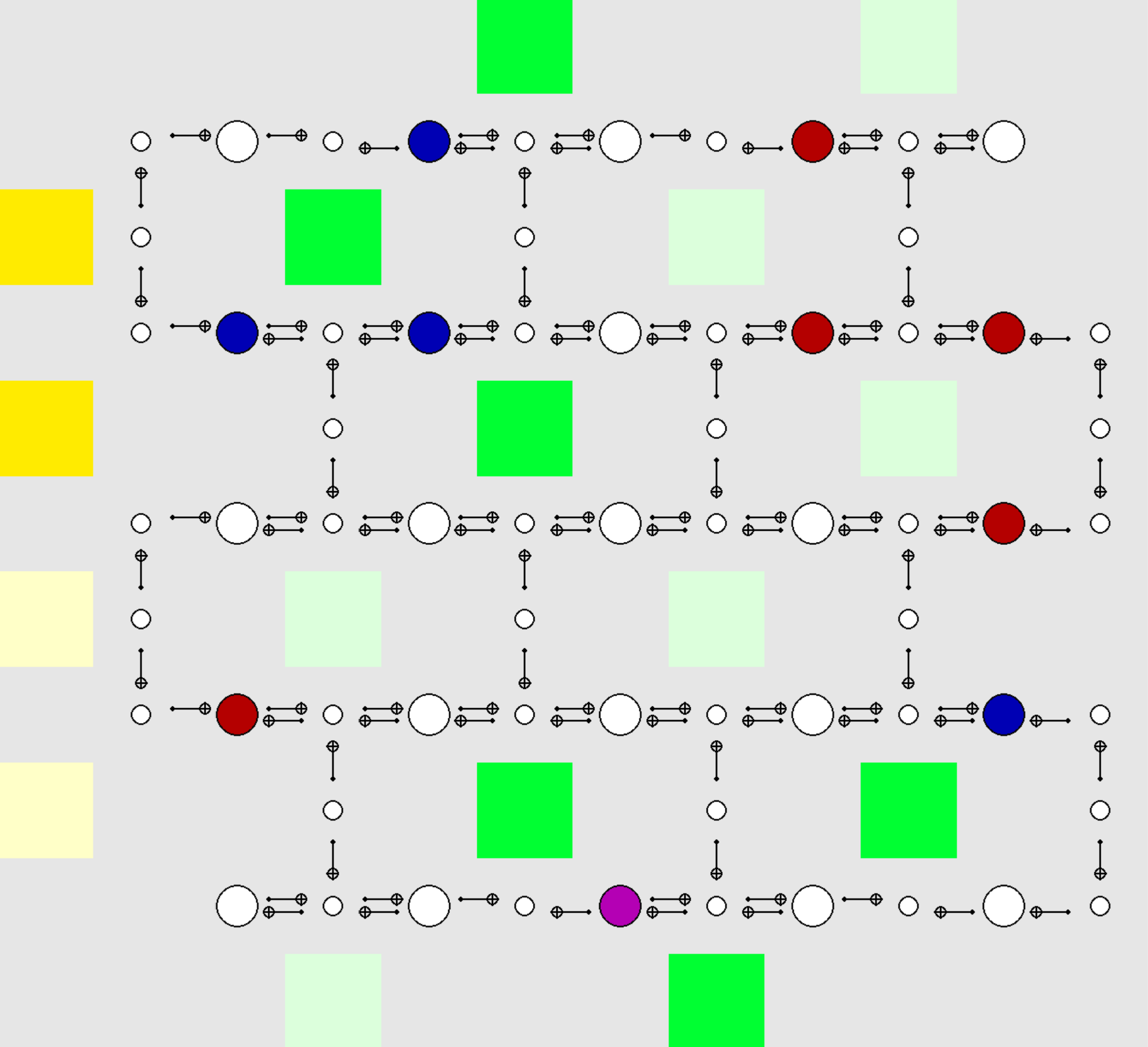}
    \caption{\textbf{Heavy Hexagon Errors on the Adjusted Lattice.} A $d=5$ example of the adjusted HH code architecture which directly corresponds to physically realised IBM quantum processors, with some data errors and related syndrome. White circles correspond to the qubits, with larger and smaller representing data and ancilla respectively. Green and wellow squares refer to the parity measurement of the $Z$ and $X$ stabilisers in their local zones, respectively. These are dull when measuring a $+1$ eigenvalue and bright when measuring a $-1$ eigenvalue. Blue circles show an $X$ error on a given qubit and red circles show a $Z$ error on a given qubit. CNOTs are drawn on the lattice with corresponding directions for targets and controls.}
    \label{fig:HHd5error}
\end{figure}

In Figure \ref{fig:HHd5error}, the yellow and green stabilisers are shown for illustrative purposes. Examples of data qubit errors are shown and the corresponding stabilisers are `lit up' from dull to bright, via the measurement of the gauge operators. The eigenvalues associated with the stabiliser operators in Figure \ref{fig:HHd5error} are:
\begin{equation}
    \begin{bmatrix}
        -1 & +1 & -1 & +1 & -1 & +1 & +1 & +1 & -1 & -1 & +1 & -1
    \end{bmatrix}
\end{equation}
\vspace{-25pt}
\begin{equation*}
    \begin{bmatrix}
        -1 & -1 & +1 & +1
    \end{bmatrix}
\end{equation*}
corresponding to the $Z$ and $X$ operator respectively. These are simplified to:
\begin{equation}\label{Syndrome_Simpification}
    \begin{bmatrix}
        1 & 0 & 1 & 0 & 1 & 0 & 0 & 0 & 1 & 1 & 0 & 1
    \end{bmatrix}
\end{equation}
\vspace{-25pt}
\begin{equation*}
    \begin{bmatrix}
        1 & 1 & 0 & 0
    \end{bmatrix}
\end{equation*}
for ease of ANN training. In Equation \ref{Syndrome_Simpification}, a zero is given where no change has occurred and 1 given when a stabiliser change has occurred; $-1$ eigenvalue to $+1$ eigenvalue. When multiple errors occur within the same parity measurement of a single stabiliser, it may have its eigenvalue inverted twice, returning to its original state. Therefore, only stabilisers, at the end of chains have their values changed, as illustrated in Figure \ref{fig:HHd5error}. This is less obvious for the $Z$ errors, as the nature of the Bacon-Shor stabiliser allows for chains to be continued anywhere across entire columns.

Errors across the lattice which are of the same form as the stabiliser elements, generators or otherwise, commute with all stabiliser generators and hence do not change the underlying information in the lattice. 
This means that the encoded state of the lattice may only be affected by a global phase and encoded information is unaltered. 
Given that the state is unaltered, gates of the same kind can be applied to the lattice wherever required, to correct for errors. 
This can be used to create sets of equivalent error chains from the same start and end points on the lattice.

\section{Comparing MWPM Results}\label{sec:MWPM}
The threshold value of 0.0045 found by Chamberland \textit{et al.} was not based on a MWPM implementation from PyMatching \cite{higgott_pymatching_2022}.
Therefore, we have implemented small distances of the HH code as described by Chamberland \textit{et al.} using PyMatching to confirm the threshold value to later confirm with the threshold value of the adjusted HH structure.
In Figure \ref{fig:HHMWPM}, the threshold of 0.005 is given for the $X$ logical error rate, which can be compared to the one given by Chamberland \textit{et al.} of 0.0045 \cite{chamberland_topological_2020}. 
These values are very similar and hence the value of 0.005 was used for comparison in the main text.

\section{Sub-Graph Locations}\label{sec:sub-graph}

\begin{figure}[t]
    \includegraphics[width=0.95\linewidth]{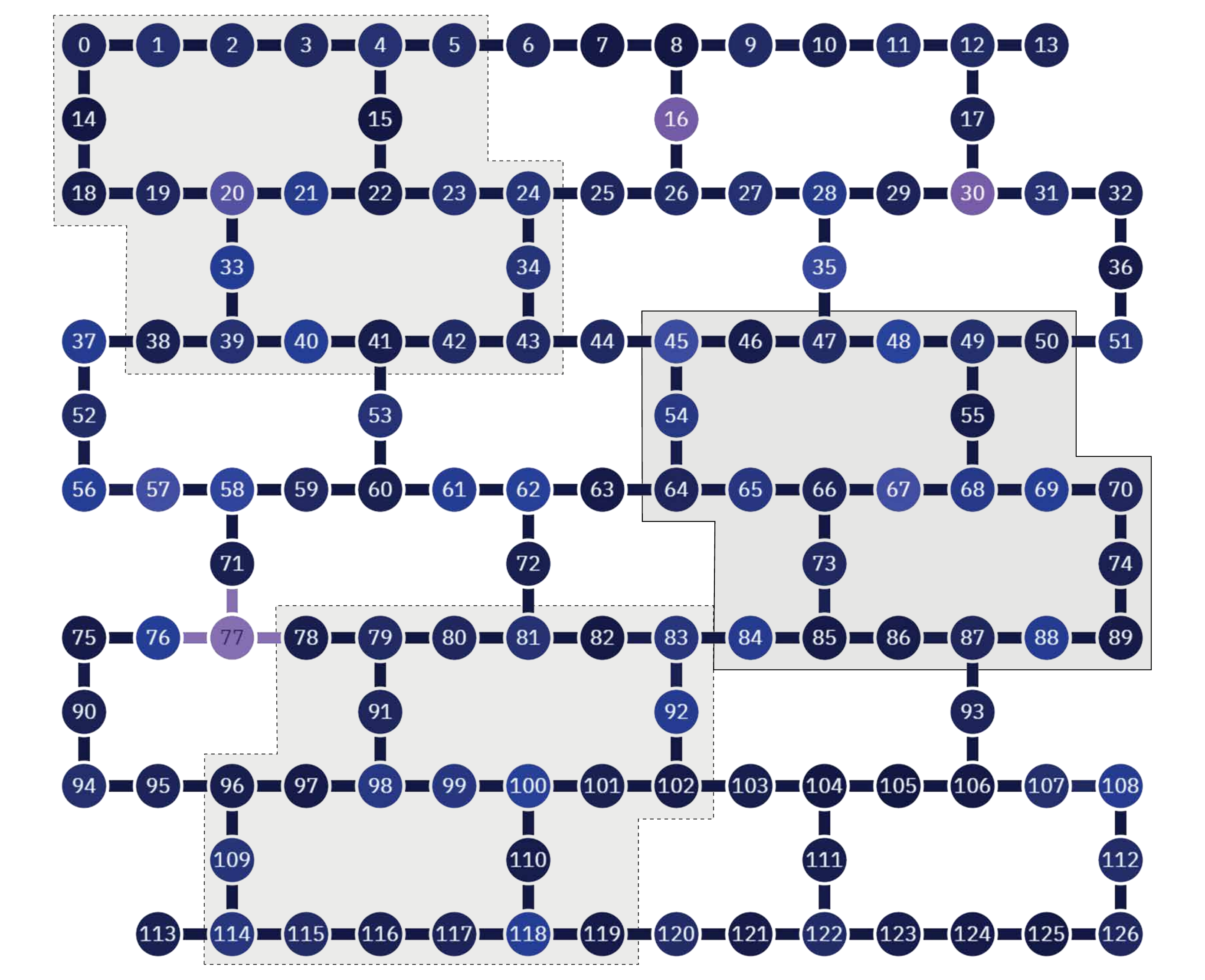}
    \centering
    \caption{\textbf{Sub-graph locations on an IBM device.} This device has 127 qubits, and therefore could support a $d=7$ HH code. A sub-graph location for a $d=3$ code is chosen. Colours correspond to error rates, with dark blue to light purple being low to high. Dashed zones indicate sub-graph locations, with the solid zone having the lowest overall error rate.}
    \label{fig:sub-graph}
\end{figure}

The five devices which were tested are capable of sustaining more than $d=5$.
\textit{Brisbane, Cusco, Nazca} and \textit{Sherbrooke} all contain 127 qubits and therefore can hold $d=7$, and \textit{Seattle} has 433 qubits and holds $d=13$.
Consequently, the position of smaller sub-graphs on these devices is important, since each qubit and pair of connected qubits are no longer uniform in their physical error probability. 
As illustrated in Figure \ref{fig:sub-graph}, different sub-graph locations have different error rates, shown by the different qubit and connection colours.
Given the variability of each sub-graph location, we constructed a systematic heuristic test which aimed to capture the overall suitability of each sub-graph location within the lattice of qubits. 
Using MWPM, each possible location was tested with only one simulated error source at a time. 
We were able to show how the logical error rate would be affected by each source.
The results from this experiment were averaged and are shown in Table \ref{tab:Badness}.

\begin{table}[h]
    \centering
    \begin{tabular}{c|c}
        Error Source & Influence Weighting\\\hline
        Single Qubit Gate Error & 1 \\
        State Initialisation Error & 17\\
        Idle Qubit Error & 41\\
        Measurement Readout Error & 65\\
        Two Qubit Gate Error & 100
    \end{tabular}
    \caption{\textbf{Heuristic Sub-graph Ranking Scores.} The logical error influence of each individual physical error source, given the same physical error rate.}
    \label{tab:Badness}
\end{table}

This table describes how the logical error rate would be affected by each physical error source, given the same underlying physical error rate, $p$.
Using this test, we were able to heuristically rank every possible sub-graph location within each devices structure.
Note that this test was completed with simulated noise only, and is not directly or explicitly related to the physically realised devices.
For the plots generated in the main text, the rankings were used to create the horizontal uncertainty for circle device markings.
The lowest ranked sub-graph's average physical error rate was used as the lower uncertainty value and highest ranked sub-graph's average physical error rate as the upper. The marking was placed on the median sub-graph's average error rate.

\end{document}